%% file: manuscript.tex
\pdfoutput=1
\documentclass[11pt,a4paper]{article}
\usepackage[utf8]{inputenc}
\usepackage[T1]{fontenc}
\usepackage{lmodern}
\usepackage{amsmath, amssymb, amsfonts}
\usepackage{graphicx}
\usepackage{booktabs}
\usepackage{multirow}
\usepackage{array}
\usepackage{xcolor}
\usepackage{hyperref}
\usepackage[authoryear,round]{natbib}
\usepackage{geometry}
\usepackage{caption}
\usepackage{float}
\usepackage[expansion=false,nopatch=footnote]{microtype}
\hypersetup{
    colorlinks=true,
    linkcolor=blue!60!black,
    citecolor=blue!60!black,
    urlcolor=blue!60!black
}

\title{Compiling and forecasting implied volatility surfaces via survey sampling, tempered inverse-probability sequence learning, and causal activity clocks}

\author{
  Deepak Bastola\textsuperscript{a}\thanks{Corresponding author: \href{mailto:dxbastola@salisbury.edu}{dxbastola@salisbury.edu}} \and
  Eli Prushansky\textsuperscript{a}\thanks{Second author: \href{mailto:eprushansky1@salisbury.edu}{eprushansky1@salisbury.edu}} \\[4pt]
  \small \textsuperscript{a}Department of Mathematical Sciences, Salisbury University, Salisbury, MD 21801, USA
}
\date{\today}

\begin{document}

\maketitle

\begin{abstract}
Compiling accurate implied volatility surfaces and forecasting derivative risk from high-frequency equity and options feeds requires addressing extreme contract heterogeneity, cross-vendor schema shifts, volatility clustering, and non-stationary trading activity. Using a multi-resolution, split-adjusted NVIDIA (NVDA) equity and options panel from 2018 through 2025, this study integrates finite-population survey sampling, tempered inverse-probability sequence learning, and causally recalibrated activity clocks. Four empirical findings emerge. First, cross-sectional surface accuracy depends critically on matching the sampling design to the target surface functional: stratified plug-in allocations using lagged cell dispersion substantially improve neutral equal-weighted surface reconstruction with complete grid coverage, whereas liquidity-weighted probability proportional to size (PPS) sampling excels on liquidity-weighted targets by construction yet inflates variance when misapplied across objectives. Second, in longitudinal sequence training of Encoder-Decoder LSTM networks, informative PPS and Neyman designs retain substantially more tail-volatility windows than simple random sampling (SRS); all three designs preserve full-census forecasting accuracy within chosen equivalence margins while cutting training time, though incremental forecasting gains of informative selection over SRS remain unestablished. Third, out-of-sample evaluation reveals a predictability contrast: short-horizon prices are not forecastable beyond a random walk with drift among tested models, whereas forward implied volatility exhibits descriptive mean-reversion skill concentrated in post-shock reversals rather than calm sessions, with parsimonious sequence models not outperforming the simpler linear HAR-IV benchmark and feature pruning driving representation stability. Fourth, across a nearly twentyfold trough-to-peak share-price expansion, rolling volume and dollar clocks stabilize bar formation rates and substantially compress intraday return tail kurtosis beyond deterministic time-of-day scaling.
\end{abstract}


\section{Introduction}
\label{sec:intro}

Quantitative modeling of equity and derivatives markets requires compiling accurate forward-looking risk indexes from noisy option chains while accounting for uneven information arrival across time \citep{black1973pricing, cont2001empirical, andersen2000benchmarking}. Over the past decade, deep recurrent architectures, including Long Short-Term Memory (LSTM) networks \citep{hochreiter1997long} and sequence-to-sequence Encoder-Decoder models \citep{sutskever2014sequence}, have been widely adopted alongside classical econometric benchmarks adapted from the Heterogeneous Autoregressive (HAR) volatility family \citep{corsi2009simple} to model financial sequences and asset return volatility \citep{sirignano2019universal, cao2019financial, liu2019deep}.

Despite rapid advances in financial data engineering and computational capacity, three methodological questions remain central to empirical derivatives research. First, how should a practitioner subsample cross-sectional option chains when compiling and reconstructing implied volatility surfaces $\sigma_{\mathrm{imp}}(m, \tau; t)$ across moneyness $m = K/S_t$ and time-to-maturity $\tau = T - t$ \citep{gatheral2004parsimonious, zhang2019deep, de2023noarbitrage, gong2022encoding}? At any market snapshot, an equity option chain mixes heavily traded, tight-spread near-the-money contracts with sparse, wide-spread out-of-the-money (OTM) wings. Classical finite-population survey sampling theory \citep{madow1949theory, horvitz1952generalization, neyman1934two, kish1965survey, cochran1977sampling, deville1998unequal, chauvet2012on} provides rigorous probability designs, specifically Probability Proportional to Size (PPS) sampling and stratified allocation, for selecting contract subsets. However, empirical options studies rarely distinguish whether the target volatility index is compiled as an equal-cell unweighted surface (via the \citealp{hajek1971comment} sample-normalized estimator) or a microstructure liquidity-weighted surface (via the \citealp{horvitz1952generalization} ratio estimator), nor whether causal plug-in stratified allocations using lagged cell dispersion are tailored to the population-weighted mean ($n_h \propto N_{h,t} S_{h,t-1}$) or to unweighted equal-cell grid accuracy ($n_h \propto S_{h,t-1}$).

Second, when training sequence-to-sequence neural networks on multi-year intraday financial panels, uniform chronological sampling devotes most gradient updates to quiet midday trading bars while sampling fewer tail-risk windows around corporate earnings releases and macroeconomic shocks \citep{lopez2018advances}. Can probability subsampling paired with tempered inverse-probability loss weighting ($d_i \propto \pi_i^{-\gamma_{\mathrm{HT}}}$) reduce computational training cost while preserving out-of-sample volatility risk forecasting relative to unweighted full-census training under formal Two One-Sided Tests (TOST) equivalence \citep{schuirmann1987comparison, lakens2017equivalence}, and does informative longitudinal selection improve out-of-sample accuracy beyond Simple Random Sampling (SRS) at the same sample budget? Moreover, when models are evaluated strictly out-of-sample without look-ahead leakage, where does multi-horizon predictability reside: in short-horizon split-adjusted equity price changes, or in mean-reverting option-implied volatility trajectories \citep{goncalves2006predictable, bernales2014can}, and how does a sequence-to-sequence LSTM compare directly against a parsimonious linear HAR-IV benchmark when evaluated with session-clustered batch-means inference \citep{diebold1995comparing, flegal2010batch, liu2022lugsail, flegal2025mcmcse}?

Third, at the high-frequency market-microstructure level, subordinated activity clocks (Volume Bars and Dollar Bars; \citealp{clark1973subordinated, ane2000order, lopez2018advances}) sample market trajectories on an activity-arrival clock rather than a chronological clock. When an underlying equity experiences a nearly twentyfold ($18.9\times$) trough-to-peak split-adjusted share-price expansion ($6.3\times$ start-to-end) over a four-year window, static activity thresholds break down. Testing whether causally recalibrated rolling volume and dollar thresholds maintain stable average bar formation rates, thin unconditional within-session return tails (Fisher excess kurtosis) beyond deterministic time-of-day diurnal volatility scaling, and alter conditional volatility clustering or return predictability requires paired session-block bootstrap inference over identical intraday trading intervals \citep{kunsch1989jackknife, efron1994introduction}.

This article addresses these three questions through a causal empirical and Monte Carlo resampling study on granular NVIDIA Corporation (NVDA) equity and options data from May 2018 through November 2025 ($40\times$ cumulative split factor across the 4-for-1 split in July 2021 and 10-for-1 split in June 2024). In cross-sectional volatility surface compilation across $432{,}529$ filtered OTM option contract records over $753$ trading days on a $4 \text{ moneyness} \times 3 \text{ maturity}$ stratification grid, Monte Carlo resampling with $1{,}000$-draw whole-day cluster bootstrap inference demonstrates that plug-in equal-cell allocation ($n_h \propto S_{h,t-1}$) and plug-in Neyman allocation ($n_h \propto N_{h,t} S_{h,t-1}$) with bounded water-filling budget rounding consistently outperform SRS across all retention fractions $\alpha \in \{0.25, 0.40, 0.60, 0.80\}$ when compiling a neutral equal-weighted surface (reducing grid RMSE at $\alpha = 0.25$ from $124.0\text{ bps}$ under SRS to $84.8\text{ bps}$ under plug-in equal-cell allocation, ratio $0.68$ $[0.66, 0.70]$, and $94.9\text{ bps}$ under plug-in Neyman, ratio $0.77$ $[0.74, 0.79]$, and at $\alpha = 0.80$ from $32.0\text{ bps}$ to $15.1\text{ bps}$, ratio $0.47$ $[0.45, 0.49]$, and $22.2\text{ bps}$, ratio $0.69$ $[0.66, 0.73]$) while guaranteeing $100\%$ stratum coverage. Conversely, Liquidity PPS sampling reduces surface compilation RMSE by $54\%\text{--}95\%$ when estimating a liquidity-weighted surface (RMSE ratios of $0.46, 0.28, 0.14,$ and $0.05$ across $\alpha \in \{0.25, 0.40, 0.60, 0.80\}$, reaching $11.7\text{ bps}$ at $\alpha=0.60$), as expected by construction when inclusion probabilities $\pi_{i,t} \propto w_{i,t}$ match the target weights, whereas applying Liquidity PPS to an equal-weighted target inflates RMSE by $2.47\text{--}3.90\times$ relative to SRS ($3.91\times$ from rounded table entries $125.1/32.0$).

In longitudinal sequence estimation across $N_{\mathrm{train}} = 9{,}139$ hourly training windows, we separate three distinct empirical findings. First, on sampling diagnostics, informative Liquidity PPS and Neyman stratified designs capture $81.2\%$ and $73.7\%$ of top-decile realized volatility ($\mathrm{RV}_{21\mathrm{h}}$) bursts at $\alpha = 0.60$ ($64.4\%$ and $48.7\%$ at $\alpha = 0.40$), compared to $59.9\%$ and $40.7\%$ under SRS, while tempered inverse-probability weights ($\gamma_{\mathrm{HT}} = 0.55$) keep Kish weight dispersion tightly bounded ($\mathrm{DEFF} \in [1.016, 1.091]$). Second, relative to full-census training on the 31-feature Tier 5a specification ($D=31$, all features plus OptionVAE latents), paired 10-seed TOST comparisons on the Encoder-Decoder LSTM under chosen equivalence margins ($\pm 0.50\%$ price RMSE and $\pm 2.00\%$ IV RMSE) show that subsampling at $\alpha = 0.60$ cuts training wall-clock time by $21\%\text{--}22\%$ ($29\%\text{--}37\%$ at $\alpha = 0.40$) and preserves out-of-sample accuracy across all three probability designs, including SRS: multi-horizon price RMSE stays within $+0.11\%\text{ to }+0.14\%$ of the full census ($p_{\mathrm{TOST}} \le 0.002$), mean single-model IV RMSE across the 10 seeds matches the full census ($1{,}377\text{ bps}$) within the $\pm 2.00\%$ margin ($p_{\mathrm{TOST}} \le 0.020$, two-sided $p_{\mathrm{diff}} \in [0.642, 0.860]$), and the 10-seed Neyman ensemble achieves $1{,}339\text{ bps}$ ($+9.9\%$ MSE skill vs.\ $1{,}346\text{ bps}$, $+8.9\%$ for the full census). Third, comparing informative subsampling directly against SRS at equal budget ($\alpha = 0.60$) shows that incremental out-of-sample forecasting gains over SRS are not established (paired 10-seed differences vs.\ SRS are $+0.02\%$ [$p=0.893$] for PPS and $+0.01\%$ [$p=0.963$] for Neyman on price RMSE, and $+0.51\%$ [$p=0.608$] and $+0.09\%$ [$p=0.937$] on IV RMSE).

Furthermore, on the unseen 2024--2025 test partition ($N_{\mathrm{test}} = 1{,}963$ hourly windows across $281$ trading sessions, or $27{,}482$ anchor-horizon pairs, $0\%$ of which target the already-known lagged $\mathrm{IV}_{d-1}$), 1-to-14 hour split-adjusted NVDA equity prices are not forecastable beyond a random walk with drift among the tested models (Random Walk RMSE $\$4.148$; Historical Drift $\$4.141$), whereas forward end-of-day implied volatility exhibits descriptive mean-reversion skill concentrated in post-shock volatility reversals: our linear HAR-IV adaptation achieves $+21.0\%$ pooled MSE skill ($1{,}254\text{ bps}$ vs.\ $1{,}410\text{ bps}$ persistence; $+22.0\%$ on strictly future-session targets and $+17.4\%\text{--}22.9\%$ in pure daily 1- and 2-session-ahead regressions), and the Encoder-Decoder LSTM trained on parsimonious Tier 2 features (IV, skew, $\Delta\mathrm{IV}_{1\mathrm{d}}$, and VRP proxy) under Liquidity PPS ($\alpha = 0.60$) achieves $+17.2\%$ ensemble MSE skill ($1{,}283\text{ bps}$; $+18.2\%$ on strictly future-session targets). Notably, the neural sequence model does not outperform the simpler linear HAR-IV benchmark ($1{,}283\text{ bps}$ vs.\ $1{,}254\text{ bps}$; direct session-clustered Diebold-Mariano test with \citealp{flegal2010batch, flegal2025mcmcse} \texttt{mcmcse} batch-means standard errors $t_{\mathrm{sess}} = +0.51, p = 0.61$). Moreover, these pooled MSE reductions are driven by a small number of post-earnings and macro tariff volatility reversals (the top 5 sessions supply $144\%\text{--}157\%$ of net squared-error reductions), do not reach statistical significance against persistence under session-clustered DM tests ($p = 0.34$ for HAR-IV, $p = 0.24\text{--}0.35$ for Tier 2 LSTMs), and coexist with a positive sample median daily squared-error differential where persistence achieves lower squared error on $57.3\%$ ($95\%$ moving-block bootstrap CI $[50.2\%, 64.8\%]$) of daily 15:00 ET anchors for HAR-IV and $53.0\%$ ($[47.0\%, 59.4\%]$) for the Tier 2 PPS LSTM. Reconstructing the 2024--2025 option summary index from full-chain OPRA.PILLAR consolidated best-bid-and-offer (cbbo-1m) end-of-day snapshots across all $466$ active trading sessions eliminates a legacy $75\text{-DTE}$ truncation break, and our representation ablation demonstrates that adding 8 noisy option-activity aggregates (Tier 3, $D=28$) degrades 5-seed mean IV skill from $+13.4\%$ (Tier 2, $D=20$) to $+3.9\%$ (census; $1{,}344\text{ bps}$, $+9.2\%$ ensemble skill) and $+1.8\%$ (PPS), whereas pruning them (Tier 2) or replacing them with 3 PCA or OptionVAE latents ($D = 23$) restores $+12.3\%$ and $+13.6\%$ skill ($p_{\mathrm{Welch}} \le 0.008$ vs.\ Tier 3; indistinguishable from Tier 2 at $p_{\mathrm{Welch}} \ge 0.61$, while OptionVAE replacement yields a descriptive 5-seed cross-seed SD under PPS of $2.8\%$ vs.\ $10.6\%$). Finally, across $376{,}838$ 1-minute NVDA bars ($375{,}866$ under strict 13:00 ET early-close truncation) over $967$ sessions ($2022\text{--}2025$), causally recalibrating volume and dollar thresholds via a trailing 60-session rolling mean maintains stable average bar formation rates ($\approx 8.5\text{--}9.0\text{ bars/session}$) across a $6.3\times$ start-to-end and $18.9\times$ trough-to-peak share-price expansion ($3.0\times\text{--}6.4\times$ expansion in rolling dollar turnover per bar) and reduces interior-session return excess kurtosis to $1.83$ ($[1.05, 2.76]$) and $2.49$ ($[1.74, 3.35]$), compared to $5.00$ ($[3.20, 7.76]$) under chronological Time Bars ($\Delta\kappa = -3.17$ $[-5.99, -1.11]$ and $-2.51$ $[-4.86, -0.75]$) and $5.09$ ($[2.48, 9.44]$) under deterministic Diurnally Standardized Time Bars (direct paired $\Delta\kappa_{\mathrm{diurn}} = -3.26$ $[-6.88, -0.53], p=0.010$ and $-2.60$ $[-5.66, -0.40], p<0.001$), with similar reductions on full-session returns ($1.67$ and $2.28$ vs.\ $14.00$ chronological and $5.22$ diurnally standardized; $\Delta\kappa_{\mathrm{diurn}} = -3.55$ $[-7.40, -0.73], p=0.002$ and $-2.94$ $[-6.28, -0.55], p<0.001$) as well as on the strictly out-of-sample 291-session holdout ($2.34$ and $4.14$ vs.\ $42.27$ chronological and $14.41$ diurnally standardized on full-session returns; $2.68$ and $4.63$ vs.\ $10.81$ and $13.50$ on interior returns).

The remainder of the article is organized as follows. Section~\ref{sec:data} describes the multi-resolution NVDA dataset, split adjustments, exact window waterfall, cross-vendor OPRA index harmonization, and latent option-state compression. Section~\ref{sec:sampling} formalizes the finite-population survey sampling designs, tempered inverse-probability weights, longitudinal sampling diagnostics, and cross-sectional option surface compilation experiments. Section~\ref{sec:models} presents the multi-horizon forecasting models, paired TOST equivalence benchmarks, and out-of-sample price and implied volatility risk forecasting results. Section~\ref{sec:bars} details the intraday bar-clock analysis, and Section~\ref{sec:conclusion} discusses implications for quantitative volatility index compilation and derivative risk management and concludes.

\section{Data Engineering and Volatility Index Harmonization}
\label{sec:data}

\subsection{Multi-Resolution Split-Adjusted NVDA Dataset and Causal Partitioning}

To eliminate artificial price discontinuities across corporate actions, all historical NVDA equity prices, per-share volumes, and option strikes are backward-adjusted to post-split equivalent terms across the 4-for-1 stock split on July 20, 2021 and the 10-for-1 stock split on June 10, 2024 (a cumulative split factor of $40.0$ prior to July 20, 2021 and $10.0$ between July 20, 2021 and June 9, 2024). Our empirical investigation integrates three synchronized data layers across longitudinal sequences, cross-sectional contract chains, and high-frequency intraday transactions.

The first layer is a longitudinal hourly equity-options panel ($2018\text{--}2025$) comprising $13{,}250$ hourly bars across $1{,}893$ trading days from May 1, 2018 to November 7, 2025 ($1{,}892$ sessions with 7 clock-hour bars stamped at 09:00--15:00 ET and $1$ session, December 24, 2019, with 6 bars; on the $16$ NYSE/Nasdaq 13:00 ET early-close holiday sessions, the Databento XNAS.ITCH hourly feed retains $47$ afternoon/late-print hourly records at 13:00--15:00 ET, which are kept for uniform 7-bar daily periodicity while the high-frequency 1-minute bar-clock study in Section~\ref{sec:bars} explicitly evaluates both uniform 09:30--15:59 ET and strict 13:00 ET early-close truncated RTH transaction timestamps). Hourly equity bars are merged with daily end-of-day (EOD) option chain summaries; because daily option chain summaries reflect end-of-session (16:00 ET) closing quotes and cumulative daily volume, all EOD option features merged onto intraday hour $t$ of trading day $d$ are lagged by one full trading session ($d-1$) so that intraday bars never access same-day closing option aggregates.

The second layer is a cross-sectional raw option chain corpus ($2019\text{--}2021$) from OptionsDX containing $432{,}529$ filtered out-of-the-money (OTM; puts with $K < S_t$, calls with $K \ge S_t$) option records across $753$ trading days (of the $757$ active NYSE/Nasdaq sessions between January 2, 2019 and December 31, 2021, with 4 vendor-omitted archive dates: June 18, 2020, November 27, 2020, December 24, 2020, and January 6, 2021; median $525$ and mean $574.4$ contracts per day) spanning moneyness $m = K/S_t \in [0.60, 1.40)$, days-to-expiration $\mathrm{DTE} \in [7, 365]$, positive bid quotes ($B_{i,t} > 0$, $A_{i,t} > B_{i,t}$), and annualized implied volatility $\sigma_{\mathrm{imp},i,t} \in (0.05, 3.00)$. The third layer is a high-frequency intraday 1-minute stream ($2022\text{--}2025$) consisting of $376{,}838$ timezone-aware (US/Eastern DST-adjusted, $09\text{:}30\text{--}15\text{:}59$ ET) 1-minute bars across $967$ trading sessions from January 3, 2022 to November 7, 2025 ($375{,}866$ strict RTH 1-minute bars when $972$ post-13:00 ET rows on the $7$ early-close holiday sessions in $2022\text{--}2025$ are truncated at 13:00 ET, within $4$ zero-trade minutes of the theoretical maximum $960 \times 390 + 7 \times 210 = 375{,}870$), over which split-adjusted NVDA share price rose $6.3\times$ start-to-end (from $\$29.82$ on January 3, 2022 to $\$188.26$ on November 7, 2025) and expanded nearly twentyfold ($18.9\times$) from its October 2022 daily-close trough of $\$11.23$ ($\$10.85$ intraday low on October 13, 2022) to its October 2025 peak of $\$212.19$.

For multi-horizon sequence modeling on the longitudinal hourly panel, each sliding window uses an input lookback of $T_{\mathrm{in}} = 21$ hourly bars (3 trading sessions) to forecast $H = 14$ hourly bars (2 trading sessions) ahead. Specifically, from the $13{,}250$ raw hourly rows, the first $139$ rows (through May 29, 2018 15:00 ET) initialize the causal 140-hour rolling feature standardization window, the next $T_{\mathrm{in}} - 1 = 20$ rows complete the first 21-bar lookback window (first valid anchor at index $159$, June 1, 2018 14:00 ET), and the final $H = 14$ rows after November 5, 2025 15:00 ET lack a complete 14-step forward target horizon, yielding $N = 13{,}250 - 159 - 14 = 13{,}077$ valid multi-horizon anchors. To prevent target leakage across split boundaries, the $13{,}077$ valid anchors are partitioned chronologically ($70\% / 15\% / 15\%$) with an $H = 14$-anchor embargo dropped at both the train-validation and validation-test boundaries ($28$ boundary anchors purged in total), yielding: (i) a training set of $N_{\mathrm{train}} = 9{,}139$ windows across $1{,}307$ trading sessions ($7 \times 1{,}307 = 9{,}149 \ge 9{,}139$), spanning June 1, 2018 (14:00 ET) to August 10, 2023 (11:00 ET); (ii) a validation set of $N_{\mathrm{val}} = 1{,}947$ windows across $279$ trading sessions ($7 \times 279 = 1{,}953 \ge 1{,}947$), spanning August 14, 2023 (12:00 ET) to September 20, 2024 (12:00 ET); and (iii) an unseen leave-out test set of $N_{\mathrm{test}} = 1{,}963$ windows across $281$ trading sessions ($7 \times 281 = 1{,}967 \ge 1{,}963$), spanning September 24, 2024 (13:00 ET) to November 5, 2025 (15:00 ET). All feature scalers and latent encoders are fit strictly on the training block.

\begin{figure}[!htbp]
    \centering
    \includegraphics[width=0.92\textwidth]{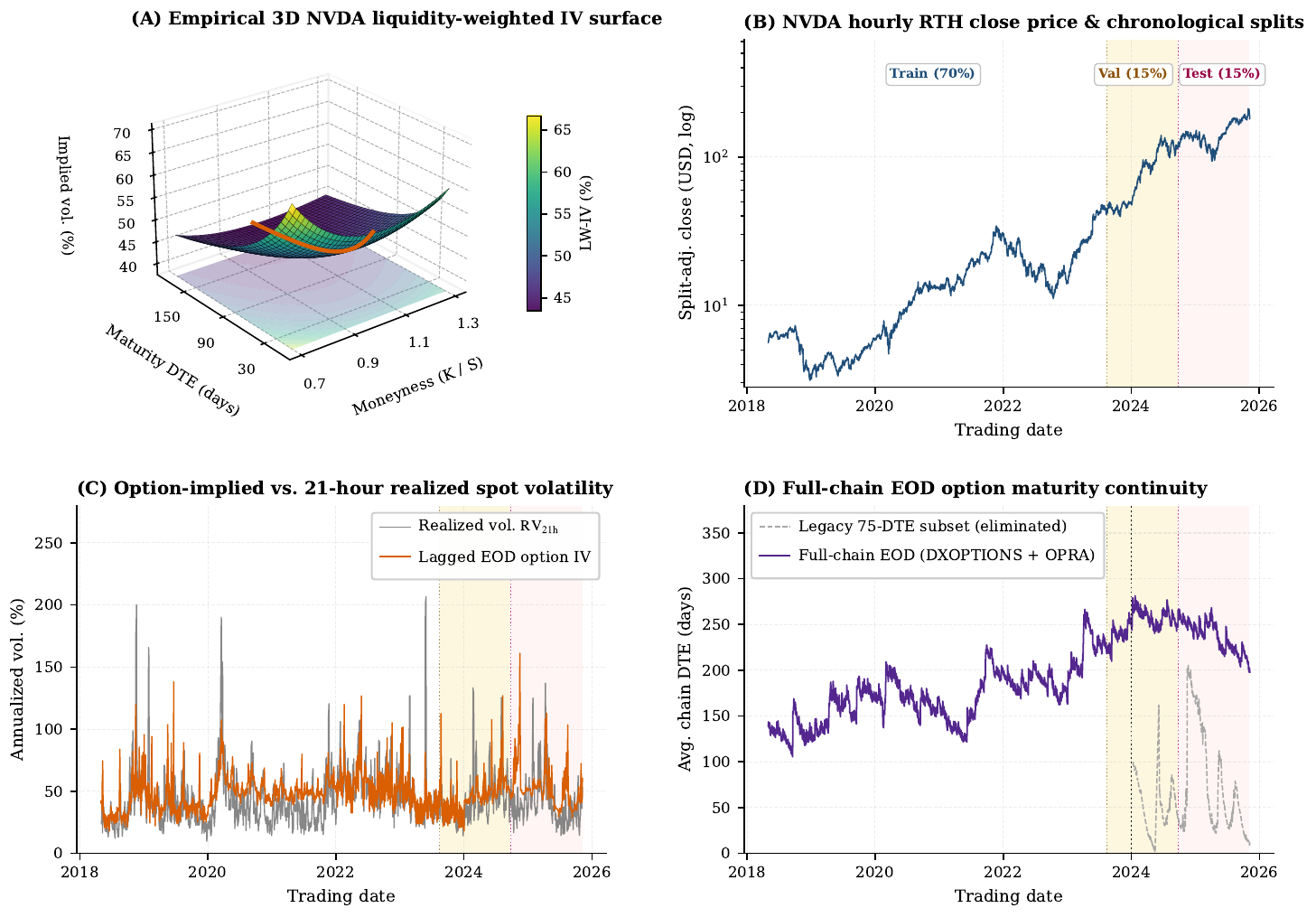}
    \caption{Granular NVDA Equity and Options Panel ($2018\text{--}2025$) and Full-Chain OPRA Index Harmonization. Panel A: Empirical 3D NVDA liquidity-weighted implied volatility surface across moneyness ($K/S \in [0.70, 1.30]$) and maturity ($\mathrm{DTE} \in [7, 180]$ days) with projected iso-volatility contours and the at-the-money ($K/S = 1.00$) term-structure backbone. Panel B: Split-adjusted hourly RTH close price (USD, log scale) across the chronological Training ($N_{\mathrm{train}} = 9{,}139$, $1{,}307$ sessions), Validation ($N_{\mathrm{val}} = 1{,}947$, $279$ sessions), and Leave-Out Test ($N_{\mathrm{test}} = 1{,}963$, $281$ sessions) partitions. Panel C: One-session-lagged end-of-day option chain implied volatility ($\mathrm{IV}$, \%) versus annualized 21-hour realized spot return volatility ($\mathrm{RV}_{21\mathrm{h}}$, \%). Panel D: Cross-vendor full-chain maturity continuity across January 1, 2024 (dashed vertical line) achieved by compiling the 2024--2025 tape from full-chain OPRA.PILLAR cbbo-1m end-of-day snapshots across all $466$ active trading sessions ($\mathrm{DTE} \approx 230\text{--}246\text{ days}$, solid curve), contrasted against the eliminated legacy $75\text{-DTE}$ filtered subset ($\sim 62\text{ days}$, grey dashed curve).}
    \label{fig:data_overview}
\end{figure}

Figure~\ref{fig:data_overview} illustrates the empirical 3D liquidity-weighted implied volatility surface (Panel A), the split-adjusted price trajectory (Panel B), the implied-versus-realized volatility dynamics (Panel C), and our cross-vendor index harmonization across the 2024 boundary (Panel D). In an initial compilation of the 2024--2025 option summary series, transitioning from pre-2024 full-chain OptionsDX records ($\sim 5{,}300$ contracts/day, mean $\mathrm{DTE} \approx 181\text{ days}$, rising to $\sim 230\text{ days}$ in 2023) to a pre-filtered near-the-money OPRA subset ($\mathrm{DTE} \le 75\text{ days}$, $\pm 15\%$ moneyness) introduced an artificial drop in average maturity to $\sim 62\text{ days}$ on January 1, 2024 (grey dashed curve in Panel D). To eliminate this index compilation break, we reconstructed the complete 2024--2025 option summary tape from full-chain OPRA.PILLAR consolidated best-bid-and-offer (cbbo-1m) snapshots at the 15:59--16:00 ET session close across all $466$ active NYSE/Nasdaq trading sessions from January 2, 2024 to November 7, 2025 ($1{,}820$ to $3{,}972$ active strike-expiration series per day across the full LEAPS horizon up to $1{,}092\text{ days}$; 5 raw vendor holiday placeholder dates are excluded upon merging with the active exchange calendar). As shown by the solid curve in Panel D, the harmonized full-chain OPRA.PILLAR compilation restores smooth maturity continuity across the 2024 boundary ($\mathrm{DTE} \approx 230\text{ days}$ in 2023 to $\approx 246\text{ days}$ in 2024--2025) and fills all 20 previously missing 2024 trading sessions.

\subsection{Nested Feature Tiers and Latent Option-State Representations: Augmentation vs.\ Replacement}

To evaluate both the incremental predictive value of option-implied risk factors and whether low-dimensional latent embeddings should augment or replace noisy option chain aggregates, we organize our hourly explanatory variables into nested and replacement representation tiers. The baseline Tier 1 specification ($D = 16$) comprises strictly causal equity OHLCV and technical indicators: 1-hour log-return, intraday high-low log range, open-to-close log return, Parkinson range volatility, annualized 7-hour, 21-hour, 35-hour, and 70-hour realized volatility ($\mathrm{RV}_{7\mathrm{h}}, \mathrm{RV}_{21\mathrm{h}}, \mathrm{RV}_{35\mathrm{h}}, \mathrm{RV}_{70\mathrm{h}}$), moving-average ratios ($\mathrm{MA}_{7/35}, \mathrm{MA}_{35/140}$), 7-hour volume surge and 35-hour dollar-volume surge, 14-hour RSI, MACD histogram, and cyclical intraday time-of-day sine/cosine encodings. Tier 2 ($D = 20$) augments Tier 1 with four core option risk metrics from session $d-1$ (standardized via a strictly causal 140-hour rolling z-score): lagged EOD chain implied volatility ($\mathrm{IV}_{t}$), put-call IV skew ($\mathrm{Skew}_{t} = \mathrm{IV}_{\mathrm{put}, t} - \mathrm{IV}_{\mathrm{call}, t}$), the 1-day IV change ($\Delta\mathrm{IV}_{1\mathrm{d},t}$), and the backward-looking variance risk premium proxy ($\mathrm{VRP}_{t}^{\mathrm{proxy}} = \mathrm{IV}_{t}^2 - \mathrm{RV}_{21\mathrm{h}, t}^2$, defined as the spread between lagged implied variance and trailing 21-hour realized spot variance). Tier 3 ($D = 28$) further incorporates eight daily option chain activity and structure aggregates: log aggregate option liquidity, log total option volume, volume-weighted average moneyness, average traded DTE, OTM contract share, ATM contract share, short-maturity ($\mathrm{DTE} < 30$) share, and long-maturity ($\mathrm{DTE} > 90$) share.

Because the eight auxiliary option activity aggregates in Tier 3 exhibit substantial cross-correlation, we fit two unsupervised 3-dimensional latent compression models strictly on the training-split daily option-state vector $\tilde{\mathbf{s}}_t \in \mathbb{R}^{10}$ (standardized via a causal 20-session rolling z-score): orthogonal Principal Component Analysis (PCA; \citealp{hotelling1933analysis}, where the top 3 components explain $23.81\%, 18.29\%,$ and $17.34\%$ of training variance, or $59.44\%$ cumulatively) and a $\beta$-Variational Autoencoder (OptionVAE; \citealp{kingma2014autoencoding, higgins2017betavae, gong2022encoding}) trained under the penalized evidence lower bound objective:
\begin{equation}
    \mathcal{L}_{\mathrm{VAE}}(\theta, \phi; \tilde{\mathbf{s}}_t) = \mathbb{E}_{q_{\phi}(\mathbf{z}_t \mid \tilde{\mathbf{s}}_t)}\!\left[\left\| \tilde{\mathbf{s}}_t - \hat{\mathbf{s}}_{\theta}(\mathbf{z}_t) \right\|_2^2\right] + \beta \, D_{\mathrm{KL}}\!\left(q_{\phi}(\mathbf{z}_t \mid \tilde{\mathbf{s}}_t) \,\|\, \mathcal{N}(\mathbf{0}, \mathbf{I}_3)\right), \quad \beta = 0.05.
    \label{eq:vae_loss}
\end{equation}
Here $\beta = 0.05 < 1$ down-weights the Kullback-Leibler divergence penalty relative to a standard VAE to preserve reconstruction fidelity on the 10-variable option state rather than enforcing strong factor disentanglement ($\beta > 1$ in \citealp{higgins2017betavae}); accordingly, only the PCA coordinates are strictly orthogonal while OptionVAE provides a regularized nonlinear embedding. Crucially, we evaluate each 3-dimensional latent embedding under two distinct architectural protocols: (i) an \emph{augmentation} protocol (Tier 4a for PCA and Tier 5a for OptionVAE, $D = 31$), which appends the 3 latent coordinates to all 28 features of Tier 3; and (ii) a \emph{replacement} protocol (Tier 4b for PCA and Tier 5b for OptionVAE, $D = 23$), which retains the 20 core price and level/skew/VRP features of Tier 2 and \emph{replaces} the 8 noisy option activity aggregates with the 3 PCA or OptionVAE latents.

Motivated by activity-based sampling concepts in \citet{lopez2018advances}, we also define a strictly causal composite Options Information Intensity score $I_t$ for this study at each hourly training anchor $t$:
\begin{equation}
    I_t = \log\!\left(1 + \mathrm{OptVol}_{t}\right) \cdot \mathrm{IV}_{t} \cdot \left(1 + \frac{|\mathrm{Skew}_t|}{\overline{|\mathrm{Skew}|_{\mathrm{train}}}}\right),
    \label{eq:info_intensity}
\end{equation}
which elevates inclusion probabilities during sessions with heavy options turnover, elevated implied volatility, and pronounced put-call skew asymmetry.

\section{Finite-Population Survey Sampling Designs and Surface Compilation}
\label{sec:sampling}

\subsection{Formalization of SRS, Madow Systematic PPS, Stratified Allocations, and Tempered Weights}

Let $\mathcal{U} = \{1, 2, \dots, N\}$ denote a finite population of either longitudinal training sequence anchors ($N = N_{\mathrm{train}} = 9{,}139$) or cross-sectional daily option contracts on trading day $t$ ($N = N_t$). Given a target retention fraction $\alpha \in (0, 1)$, we select a sample $\mathcal{S} \subset \mathcal{U}$ of fixed integer size $n = \mathrm{round}(\alpha N)$ without replacement under probability sampling designs \citep{madow1949theory, horvitz1952generalization, neyman1934two, cochran1977sampling, deville1998unequal}. Under Simple Random Sampling (SRS), every unit $i \in \mathcal{U}$ has an equal first-order inclusion probability $\pi_i^{\mathrm{SRS}} = n / N$ and unit design weight $d_i = 1$.

Under Liquidity-Weighted Probability Proportional to Size (PPS), each unit $i \in \mathcal{U}$ is assigned a strictly positive size measure $M_i > 0$, given by the information intensity $I_i$ in Equation~\eqref{eq:info_intensity} for longitudinal sequence anchors or the contract microstructure liquidity score
\begin{equation}
    w_{i,t} = \frac{\mathrm{Vol}_{i,t} + 1}{\mathrm{RelSpread}_{i,t} + 10^{-4}}, \qquad \mathrm{RelSpread}_{i,t} = \frac{A_{i,t} - B_{i,t}}{0.5(A_{i,t} + B_{i,t})},
    \label{eq:contract_liq}
\end{equation}
for cross-sectional option chains. Exact first-order inclusion probabilities $\pi_i^{\mathrm{PPS}} = \min(1, c \, M_i)$ satisfying $\sum_{i=1}^{N} \pi_i^{\mathrm{PPS}} = n$ are obtained via iterative certainty capping: any unit with $n M_i / \sum_j M_j \ge 1$ is assigned $\pi_i^{\mathrm{PPS}} = 1$ (certainty inclusion), and the remaining sample budget $n - |\mathcal{U}_{\mathrm{cert}}|$ is distributed proportionally across non-certainty units until all $\pi_i^{\mathrm{PPS}} \le 1$. To draw a fixed-size without-replacement sample realizing $\pi_i^{\mathrm{PPS}}$, we employ \citet{madow1949theory} single-start randomized systematic PPS sampling on a randomly permuted frame \citep[][pp.~265--266]{cochran1977sampling}: after randomly permuting the population indices, we compute partial sums $C_k = \sum_{j=1}^{k} \pi_{(j)}^{\mathrm{PPS}}$, draw a single uniform start $U \sim \mathrm{Uniform}(0, 1)$, and select the units whose cumulative intervals $(C_{k-1}, C_k]$ contain the $n$ equispaced grid points $U + r$ for $r = 0, 1, \dots, n-1$ \citep[related fixed-size unequal-probability schemes include ordered pivotal sampling and Deville's systematic procedure, whose equivalence is analyzed in][]{deville1998unequal, chauvet2012on}. A $5{,}000$-replicate Monte Carlo verification on the longitudinal training frame ($\alpha = 0.60, n = 5{,}483$) confirms that randomized systematic PPS realizes the target inclusion probabilities with mean absolute deviation $\frac{1}{N}\sum_{i=1}^N |\hat{\pi}_i - \pi_i^{\mathrm{PPS}}| = 0.0044$ (compatible with zero pure Monte Carlo error $\sqrt{\pi_i(1-\pi_i)/5000} \le 0.007$).

Under Stratified Sampling, the population $\mathcal{U}$ is partitioned into $H = 12$ mutually exclusive strata $\mathcal{U} = \bigcup_{h=1}^{12} \mathcal{U}_h$. For longitudinal training sequences, strata are formed by the $4 \times 3$ Cartesian product of training-split quartiles of $\mathrm{IV}_t$ and terciles of $\mathrm{Skew}_t$, with within-stratum dispersion $S_h$ measured by the stratum standard deviation of $\mathrm{RV}_{21\mathrm{h},t}$. For cross-sectional option chains, strata correspond to a $4 \text{ moneyness} \times 3 \text{ maturity}$ grid with moneyness breakpoints $[0.60, 0.90), [0.90, 1.00), [1.00, 1.10), [1.10, 1.40)$ and maturity breakpoints $[7, 30), [30, 90), [90, 366)$ calendar days; to prevent any look-ahead leakage on day $t$, we replace the unknown contemporaneous stratum dispersion $S_{h,t}$ with the prior trading day's within-cell implied volatility standard deviation $S_{h,t-1}$ (initialized on day 1 by the cross-cell mean dispersion). In textbook stratified simple random sampling without replacement when true stratum standard deviations $S_h$ are known, the sampling variance of a general linear combination of stratum means $\hat{\theta} = \sum_{h=1}^H a_h \hat{\bar{Y}}_h$ is $\mathrm{Var}(\hat{\theta}) = \sum_{h=1}^H a_h^2 (1 - n_h/N_h) S_h^2 / n_h$, whose unconstrained optimum subject to $\sum_{h=1}^H n_h = n$ satisfies $n_h \propto |a_h| S_h$. Substituting lagged dispersion $S_{h,t-1}$ and applying bounded water-filling rounding therefore yields two causal plug-in stratified allocations:
\begin{equation}
    n_h^{\mathrm{Neyman}} \propto N_h S_h \quad (a_h = N_h/N), \qquad
    n_h^{\mathrm{EqCell}} \propto S_h \quad (a_h = 1/H), \qquad
    \pi_i = \frac{n_h}{N_h} \quad \text{for } i \in \mathcal{U}_h.
    \label{eq:neyman}
\end{equation}
Plug-in \citet{neyman1934two} allocation ($n_h \propto N_{h,t} S_{h,t-1}$) approximates the oracle allocation for the population-weighted total or overall mean ($a_h = N_h / N$), whereas Plug-in Equal-Cell allocation ($n_h \propto S_{h,t-1}$) approximates the oracle allocation for the unweighted average of cell estimation variances across the 12-cell $(m, \tau)$ surface grid ($a_h = 1/H$). In both designs, integer stratum sizes satisfying $1 \le n_h \le N_h$ and the exact fixed budget $\sum_{h=1}^{12} n_h = n$ are enforced via bounded water-filling and largest-remainder rounding: any stratum where the initial proportional allocation exceeds $N_h$ is capped at $n_h = N_h$ ($\pi_i = 1$), and the surplus sample budget is iteratively redistributed across unsaturated strata before rounding fractional remainders.

During longitudinal neural network training, unmodified \citet{horvitz1952generalization} inverse-probability weights $d_i^{\mathrm{HT}} = \pi_i^{-1}$ provide an unbiased estimator of the full-census empirical risk $\mathcal{R}_{\mathrm{census}}(\theta) = \frac{1}{N}\sum_{i=1}^N \ell_i(\theta)$, but can inflate stochastic mini-batch gradient variance when $\pi_i$ varies across volatility regimes. We therefore employ tempered inverse-probability weights $d_i(\gamma_{\mathrm{HT}}) = \pi_i^{-\gamma_{\mathrm{HT}}}$ and normalize them to unit sample mean $\tilde{d}_i = n \, d_i / \sum_{j \in \mathcal{S}} d_j$ in the empirical training objective via a \citet{hajek1971comment} ratio normalization \citep[][Ch.~6]{cochran1977sampling}. Prior to sample-mean normalization, the expected unnormalized tempered empirical risk over the sampling design satisfies:
\begin{equation}
    \mathbb{E}_{\mathcal{S}}\!\left[\sum_{i \in \mathcal{S}} \pi_i^{-\gamma_{\mathrm{HT}}} \ell_i(\theta)\right] = \sum_{i=1}^{N} \pi_i^{1 - \gamma_{\mathrm{HT}}} \ell_i(\theta) =: N \, \mathcal{R}_{\gamma_{\mathrm{HT}}}(\theta).
    \label{eq:tempered_risk}
\end{equation}
Equation~\eqref{eq:tempered_risk} makes the bias-variance trade-off explicit: prior to sample-mean normalization, $\gamma_{\mathrm{HT}} = 1$ recovers design-unbiased Horvitz-Thompson risk estimation ($\mathcal{R}_{1} = \mathcal{R}_{\mathrm{census}}$; after sample-mean normalization $\tilde{d}_i = n \, d_i / \sum_{j \in \mathcal{S}} d_j$, $\gamma_{\mathrm{HT}} = 1$ corresponds to self-normalized H\'{a}jek ratio inverse-probability weighting), $\gamma_{\mathrm{HT}} = 0$ corresponds to unweighted informative subsampling tilted by $\pi_i$, and our baseline $\gamma_{\mathrm{HT}} = 0.55$ targets a mildly volatility-tilted population risk ($\pi_i^{0.45}$) that removes $55\%$ of the log-inclusion tilt while tightly bounding the \citet{kish1965survey} effective sample size and weight-dispersion factor ($\mathrm{DEFF}$):
\begin{equation}
    N_{\mathrm{eff}} = \frac{\left(\sum_{i \in \mathcal{S}} d_i\right)^2}{\sum_{i \in \mathcal{S}} d_i^2}, \qquad \mathrm{DEFF} = \frac{n}{N_{\mathrm{eff}}} = 1 + \mathrm{CV}^2(d_i).
    \label{eq:deff}
\end{equation}

\subsection{Longitudinal Sequence Sampling Diagnostics}

\begin{table}[!htbp]
\centering
\caption{Longitudinal Survey Sampling Diagnostics Across Training Sequences ($N_{\mathrm{train}} = 9{,}139$ Hourly Windows, 10-Seed Mean).}
\label{tab:sampling}
{\small\setlength{\tabcolsep}{4.5pt}\IfFileExists{tables/tab_sampling.tex}{\input{tables/tab_sampling.tex}}{\input{tab_sampling.tex}}}
\vspace{2pt}
\begin{minipage}{0.98\textwidth}
\footnotesize \textit{Notes:} Compares Simple Random Sampling (SRS), Liquidity-Weighted Probability Proportional to Size (PPS), and Neyman stratified allocation across retention budgets $\alpha \in \{0.40, 0.60, 0.80\}$ on the training population ($N_{\mathrm{train}} = 9{,}139$). Columns report exact fixed sample size $n = \mathrm{round}(\alpha N_{\mathrm{train}})$ (enforced via bounded water-filling and largest-remainder rounding; without water-filling redistribution of saturated strata, simple clipping at $\alpha=0.80$ yields $n=7{,}101$ and $\mathrm{DEFF}=1.017$), Kish effective sample size $N_{\mathrm{eff}}$, Kish survey design effect $\mathrm{DEFF} = n/N_{\mathrm{eff}}$ under tempered inverse-probability weights ($\gamma_{\mathrm{HT}} = 0.55$), non-circular capture rate (\%) of top-decile realized volatility ($\mathrm{RV}_{21\mathrm{h}}$) bursts, maximum chronological gap between retained anchors (RTH hours), and the 10-seed realized sample average 1-Wasserstein distance $W_1(\mathrm{IV})$ (in bps) between the unweighted sample and full training population (at $\alpha=0.60$, the 10-seed realized sample mean IV shift and $W_1(\mathrm{IV})$ are identically $+395.5\text{ bps}$ [$395\text{ bps}$] for PPS and $+251.0\text{ bps}$ [$250\text{ bps}$] for Neyman, whereas the analytical design-expected mean shifts under $\pi_i$ in Eq.~\ref{eq:tempered_risk} are $+398.4\text{ bps}$ [$+398\text{ bps}$] and $+253.2\text{ bps}$ [$+253\text{ bps}$]).
\end{minipage}
\end{table}

\begin{figure}[!htbp]
    \centering
    \includegraphics[width=0.92\textwidth]{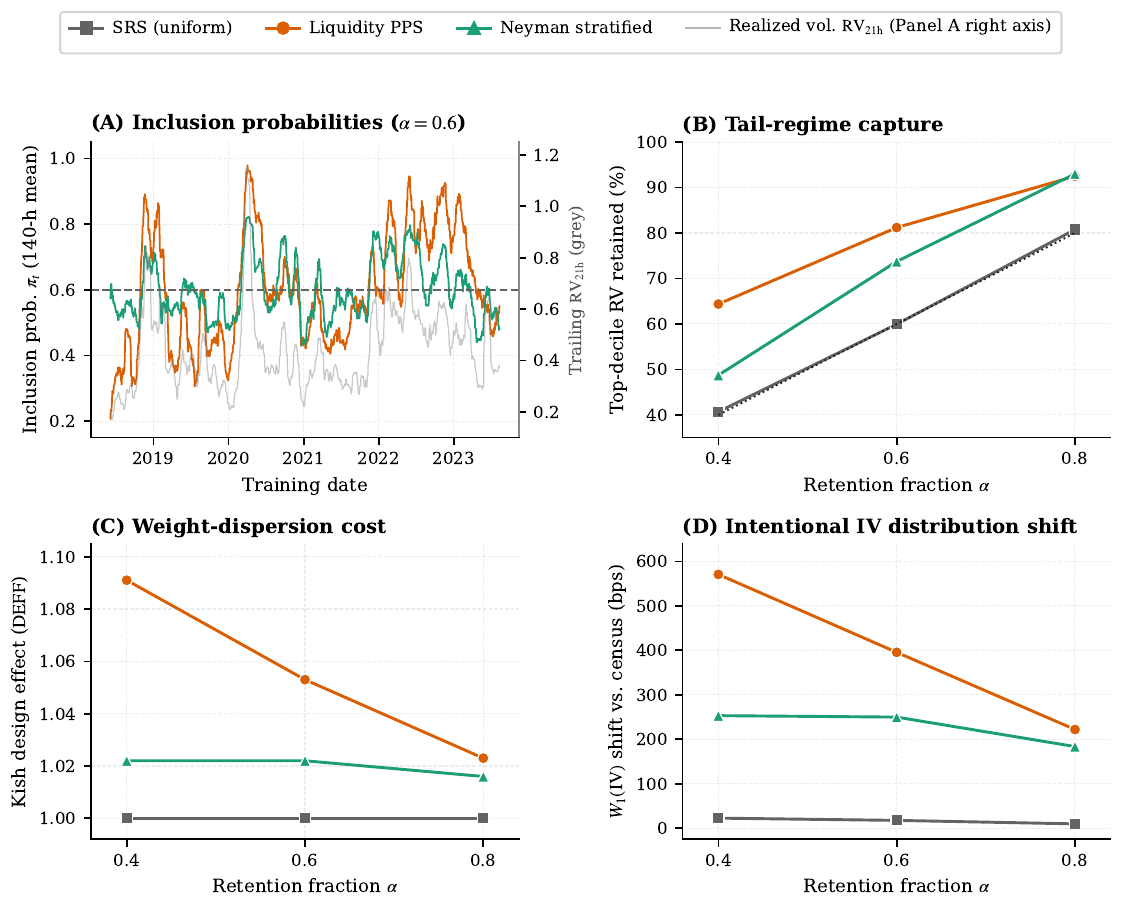}
    \caption{Longitudinal Survey Sampling Diagnostics on the Training Partition ($N_{\mathrm{train}} = 9{,}139$ Hours). Panel A: Trailing 140-hour (20-session) moving average of inclusion probabilities $\pi_t$ at $\alpha = 0.60$ for Liquidity PPS and Neyman stratified sampling overlaid against realized volatility $\mathrm{RV}_{21\mathrm{h}}$ (grey). Panel B: Percentage of top-decile $\mathrm{RV}_{21\mathrm{h}}$ tail windows retained across $\alpha \in \{0.4, 0.6, 0.8\}$ compared to the SRS expectation ($\pi = \alpha$). Panel C: Kish Survey Design Effect ($\mathrm{DEFF}$) quantifying inverse-probability weight dispersion. Panel D: Unweighted Wasserstein-1 shift $W_1(\mathrm{IV})$ (bps) demonstrating deliberate oversampling of high-volatility states.}
    \label{fig:sampling_diagnostics}
\end{figure}

Table~\ref{tab:sampling} and Figure~\ref{fig:sampling_diagnostics} report the empirical diagnostics of SRS, Liquidity PPS, and Neyman stratified sampling across the $N_{\mathrm{train}} = 9{,}139$ training sequences. Because inclusion probabilities $\pi_i$ under PPS and Neyman depend on lagged option-implied metrics ($\mathrm{IV}_t, \mathrm{Skew}_t, \mathrm{OptVol}_t$), we evaluate tail-regime capture against a strictly non-circular spot benchmark: the proportion of training windows falling in the top decile ($90\text{th}$ percentile) of realized equity return volatility $\mathrm{RV}_{21\mathrm{h},t}$.

As shown in Table~\ref{tab:sampling} and Panel B of Figure~\ref{fig:sampling_diagnostics}, both informative designs substantially oversample extreme realized-volatility regimes relative to SRS. At $\alpha = 0.40$ ($n = 3{,}656$), SRS retains $40.7\%$ of top-decile $\mathrm{RV}_{21\mathrm{h}}$ bursts, whereas Neyman stratification retains $48.7\%$ and Liquidity PPS retains $64.4\%$. At $\alpha = 0.60$ ($n = 5{,}483$), top-decile $\mathrm{RV}_{21\mathrm{h}}$ capture rises to $73.7\%$ under Neyman stratification and $81.2\%$ under Liquidity PPS (compared to $59.9\%$ under SRS), and at $\alpha = 0.80$ ($n = 7{,}311$) both informative designs capture $92.5\%\text{--}93.0\%$ of tail bursts. Simultaneously, tempered inverse-probability weighting ($\gamma_{\mathrm{HT}} = 0.55$) keeps weight dispersion tightly bounded: Kish $\mathrm{DEFF}$ is at most $1.022$ for Neyman stratification ($N_{\mathrm{eff}} = 5{,}366$ at $\alpha=0.60$) and $1.053$ for Liquidity PPS ($N_{\mathrm{eff}} = 5{,}205$ at $\alpha=0.60$). Panel D of Figure~\ref{fig:sampling_diagnostics} and Equation~\eqref{eq:tempered_risk} quantify the resulting distributional shift relative to the training census mean implied volatility ($49.67\%$): at $\alpha = 0.60$, under the analytical design expectation $\sum_{i \in \mathcal{U}} \pi_i^{1-\gamma_{\mathrm{HT}}} \mathrm{IV}_i / \sum_{i \in \mathcal{U}} \pi_i^{1-\gamma_{\mathrm{HT}}} - \overline{\mathrm{IV}}_{\mathcal{U}}$ from Equation~\eqref{eq:tempered_risk}, unweighted informative subsampling ($\gamma_{\mathrm{HT}} = 0.00$) shifts the design-expected training mean IV by $+398\text{ bps}$ under Liquidity PPS and $+253\text{ bps}$ under Neyman stratification (while across the 10 realized Monte Carlo draws in Table~\ref{tab:sampling}, the realized sample mean shift and 1-Wasserstein distance $W_1(\mathrm{IV})$ are identically $+395.5\text{ bps}$ [$395\text{ bps}$] for PPS and $+251.0\text{ bps}$ [$250\text{ bps}$] for Neyman, satisfying $W_1(F_{\mathcal{S}}, F_{\mathcal{U}}) = |\Delta\bar{\mu}|$ on every draw), tempered weighting ($\gamma_{\mathrm{HT}} = 0.55$) reduces this design-expected population tilt by more than half to $+182\text{ bps}$ (PPS) and $+112\text{ bps}$ (Neyman), and self-normalized inverse-probability weighting ($\gamma_{\mathrm{HT}} = 1.00$, H\'{a}jek ratio IPW) eliminates the population tilt entirely ($0\text{ bps}$).

\subsection{Cross-Sectional Volatility Surface Compilation: Target Alignment and Plug-In Equal-Cell Allocation}

\begin{figure}[!htbp]
    \centering
    \includegraphics[width=0.92\textwidth]{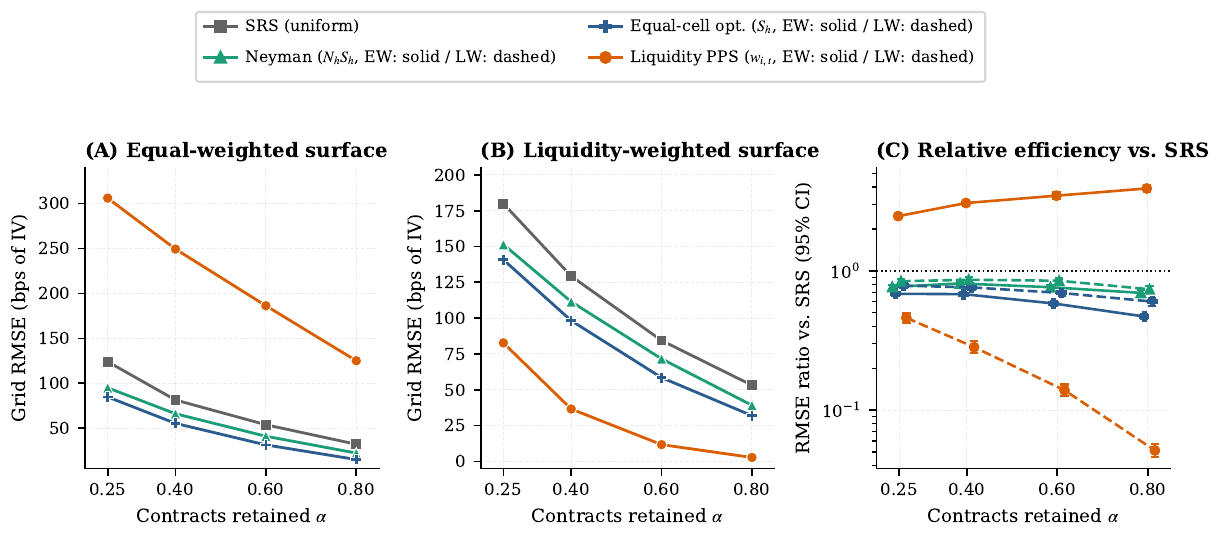}
    \caption{Cross-Sectional Option Surface Compilation Under Equal-Weighted and Liquidity-Weighted Targets ($432{,}529$ Contracts, $753$ Days). Panel A: Grid RMSE (bps of IV) when compiling the neutral Equal-Weighted (EW) cell surface. Panel B: Grid RMSE (bps of IV) when compiling the Liquidity-Weighted (LW) cell surface. Panel C: RMSE ratio relative to SRS (log scale, with $95\%$ whole-day cluster bootstrap CIs), illustrating how Plug-in Equal-Cell ($n_h \propto S_{h,t-1}$) and Plug-in Neyman ($n_h \propto N_{h,t} S_{h,t-1}$) allocations dominate on the neutral EW target across all budgets $\alpha \in [0.25, 0.80]$ while Liquidity PPS dominates on the LW target.}
    \label{fig:cross_section}
\end{figure}

We next evaluate how SRS, plug-in Neyman allocation ($n_h \propto N_{h,t} S_{h,t-1}$), plug-in Equal-Cell allocation ($n_h \propto S_{h,t-1}$), and Liquidity PPS perform when compiling and reconstructing the daily 12-cell $(m, \tau)$ implied volatility surface across $432{,}529$ filtered OTM NVDA option contracts over $753$ trading days ($2019\text{--}2021$). On each day $t$ and for each retention fraction $\alpha \in \{0.25, 0.40, 0.60, 0.80\}$ (averaged over $R = 3$ independent Monte Carlo draws per day), we draw a contract subsample $\mathcal{S}_t \subset \mathcal{U}_t$ of fixed size $n_t = \mathrm{round}(\alpha N_t)$ and compile cell $h \in \{1, \dots, 12\}$ against two distinct full-chain benchmarks. Under the neutral Equal-Weighted (EW) target surface, the full-chain benchmark in cell $h$ is the simple cell mean $\bar{\sigma}_{h,t}^{\mathrm{EW}} = \frac{1}{N_{h,t}} \sum_{i \in \mathcal{U}_{h,t}} \sigma_{i,t}$, compiled from $\mathcal{S}_t$ via the \citet{hajek1971comment} sample-normalized inverse-probability estimator:
\begin{equation}
    \hat{\sigma}_{h,t}^{\mathrm{EW}} = \frac{\sum_{i \in \mathcal{S}_{h,t}} \sigma_{i,t} / \pi_{i,t}}{\sum_{i \in \mathcal{S}_{h,t}} 1 / \pi_{i,t}}.
\end{equation}
Under the Liquidity-Weighted (LW) target surface, the full-chain benchmark in cell $h$ is the liquidity-weighted cell mean $\bar{\sigma}_{h,t}^{\mathrm{LW}} = \left(\sum_{i \in \mathcal{U}_{h,t}} w_{i,t} \sigma_{i,t}\right) / \left(\sum_{i \in \mathcal{U}_{h,t}} w_{i,t}\right)$ using $w_{i,t}$ from Equation~\eqref{eq:contract_liq}, compiled from $\mathcal{S}_t$ via the \citet{horvitz1952generalization} ratio estimator:
\begin{equation}
    \hat{\sigma}_{h,t}^{\mathrm{LW}} = \frac{\sum_{i \in \mathcal{S}_{h,t}} w_{i,t} \sigma_{i,t} / \pi_{i,t}}{\sum_{i \in \mathcal{S}_{h,t}} w_{i,t} / \pi_{i,t}}.
\end{equation}
Let $\mathcal{P}_t = \{h \in \{1,\dots,12\} : N_{h,t} > 0\}$ denote the set of non-empty population cells on day $t$; if a sampling design without stratum guarantees (SRS or PPS) leaves a cell $h \in \mathcal{P}_t$ empty in draw $r$ ($\mathcal{S}_{h,t}^{(r)} = \emptyset$), it is imputed via the global design-weighted sample mean $\hat{\bar{\sigma}}_t^{(r)} = \left(\sum_{i \in \mathcal{S}_t^{(r)}} \tilde{w}_{i,t} \sigma_{i,t}\right) / \left(\sum_{i \in \mathcal{S}_t^{(r)}} \tilde{w}_{i,t}\right)$, where $\tilde{w}_{i,t} = 1/\pi_{i,t}$ for EW and $w_{i,t}/\pi_{i,t}$ for LW. Daily surface compilation accuracy is evaluated by the equal-cell root mean squared error across present cells:
\begin{equation}
    \mathrm{RMSE}_{t}^{(r)} = \sqrt{\frac{1}{|\mathcal{P}_t|} \sum_{h \in \mathcal{P}_t} \left(\hat{\sigma}_{h,t}^{(r)} - \bar{\sigma}_{h,t}\right)^2}, \qquad
    \mathrm{RMSE}_{\mathrm{pool}} = \sqrt{\frac{1}{T R} \sum_{t=1}^{T} \sum_{r=1}^{R} \left(\mathrm{RMSE}_{t}^{(r)}\right)^2}.
    \label{eq:grid_rmse}
\end{equation}
Table~\ref{tab:cross} and Figure~\ref{fig:cross_section} report $\mathrm{RMSE}_{\mathrm{pool}}$ (in bps of IV) and the RMSE ratio relative to SRS with $95\%$ confidence intervals from a $1{,}000$-replicate whole-day cluster bootstrap \citep{efron1994introduction} that resamples entire trading days with replacement to preserve all within-day cross-sectional error dependence across cells and designs.

\begin{table}[!htbp]
\centering
\caption{Cross-Sectional Option Chain Implied Volatility Surface Compilation ($432{,}529$ Contracts Across $753$ Trading Days, $2019\text{--}2021$).}
\label{tab:cross}
\resizebox{\textwidth}{!}{\IfFileExists{tables/tab_cross.tex}{\input{tables/tab_cross.tex}}{\input{tab_cross.tex}}}
\vspace{2pt}
\begin{minipage}{0.98\textwidth}
\footnotesize \textit{Notes:} Panel A evaluates surface compilation accuracy against the neutral Equal-Weighted (EW) full-chain target surface using the Hajek inverse-probability estimator. Panel B evaluates compilation accuracy against the microstructure Liquidity-Weighted (LW) full-chain target surface using the Horvitz-Thompson ratio estimator. Plug-in Neyman allocation ($n_h \propto N_{h,t} S_{h,t-1}$) and Plug-in Equal-Cell allocation ($n_h \propto S_{h,t-1}$) use strictly causal prior-day within-cell dispersion $S_{h,t-1}$ across the $4\text{ moneyness} \times 3\text{ maturity}$ grid with bounded water-filling rounding ($\sum_h n_h = n_t$) and guarantee $100.0\%$ cell coverage at all retention fractions $\alpha$. Bracketed intervals denote $95\%$ confidence intervals from a $1{,}000$-replicate whole-day cluster bootstrap \citep{efron1994introduction}.
\end{minipage}
\end{table}

Table~\ref{tab:cross} and Figure~\ref{fig:cross_section} establish two theoretical and empirical principles for option volatility surface compilation. First, our primary cross-sectional empirical result is on the neutral Equal-Weighted (EW) surface (Panel A), where each of the 12 $(m, \tau)$ cells carries equal weight in Equation~\eqref{eq:grid_rmse} and no single design is favored by target-weight construction. Here, both causal plug-in stratified allocations with bounded water-filling budget rounding outperform SRS across all four retention budgets $\alpha \in \{0.25, 0.40, 0.60, 0.80\}$ while guaranteeing $100.0\%$ cell coverage. Consistent with Equation~\eqref{eq:neyman}, while plug-in Neyman allocation ($n_h \propto N_{h,t} S_{h,t-1}$) reduces EW grid RMSE relative to SRS by $19\%\text{--}31\%$ (from $124.0\text{ bps}$ to $94.9\text{ bps}$ at $\alpha = 0.25$, ratio $0.77$ $[0.74, 0.79]$, and from $32.0\text{ bps}$ to $22.2\text{ bps}$ at $\alpha = 0.80$, ratio $0.69$ $[0.66, 0.73]$), plug-in Equal-Cell allocation ($n_h \propto S_{h,t-1}$) achieves the lowest overall EW grid RMSE at every budget: $84.8\text{ bps}$ at $\alpha = 0.25$ (ratio $0.68$ $[0.66, 0.70]$), $55.3\text{ bps}$ at $\alpha = 0.40$ (ratio $0.68$ $[0.66, 0.70]$), $31.3\text{ bps}$ at $\alpha = 0.60$ (ratio $0.58$ $[0.56, 0.60]$), and $15.1\text{ bps}$ at $\alpha = 0.80$ (ratio $0.47$ $[0.45, 0.49]$).

Second, when the target surface is compiled as the liquidity-weighted mean $\bar{\sigma}_{h,t}^{\mathrm{LW}}$ (Panel B), choosing inclusion probabilities $\pi_{i,t}^{\mathrm{PPS}} \propto w_{i,t}$ matches the target weights by construction: it includes the highest-$w_{i,t}$ contracts with certainty ($\pi_{i,t}^{\mathrm{PPS}} = 1$, contributing zero sampling error on the dominant terms of $\bar{\sigma}_{h,t}^{\mathrm{LW}}$) and makes the Horvitz-Thompson ratio weight $w_{i,t} / \pi_{i,t}^{\mathrm{PPS}}$ constant across non-certainty contracts. As expected from this exact design-estimand alignment, Liquidity PPS reduces LW grid RMSE relative to SRS by $54\%$ at $\alpha = 0.25$ ($82.7\text{ bps}$ vs.\ $179.9\text{ bps}$, ratio $0.46$ $[0.42, 0.50]$), by $72\%$ at $\alpha = 0.40$ ($36.6\text{ bps}$ vs.\ $129.3\text{ bps}$, ratio $0.28$ $[0.26, 0.31]$), by $86\%$ at $\alpha = 0.60$ ($11.7\text{ bps}$ vs.\ $84.4\text{ bps}$, ratio $0.14$ $[0.13, 0.15]$), and by $95\%$ at $\alpha = 0.80$ ($2.7\text{ bps}$ vs.\ $53.2\text{ bps}$, ratio $0.05$ $[0.05, 0.06]$), while Plug-in Equal-Cell and Neyman allocations also reduce LW RMSE by $22\%\text{--}40\%$ and $14\%\text{--}26\%$ over SRS. Conversely, misapplying Liquidity PPS to compile the neutral EW target inflates RMSE by $2.47\text{--}3.90\times$ relative to SRS ($3.91\times$ from the rounded table entries $125.1/32.0$ at $\alpha=0.80$; $306.1\text{ bps}$ at $\alpha=0.25$ and $186.3\text{ bps}$ at $\alpha=0.60$) because rare illiquid contracts that happen to be sampled receive large inverse-probability weights $1/\pi_{i,t}^{\mathrm{PPS}}$.

To provide an illustrative analytical sensitivity calculation connecting cross-sectional compilation noise to the longitudinal EOD implied volatility forecasting task in Section~\ref{sec:models}, suppose an EOD volatility index is observed with serially uncorrelated compilation error of root mean squared magnitude $\sigma_{\mathrm{comp}}$. For a linear HAR-IV forecast $\hat{\sigma}_{t+h}^{\mathrm{HAR}} = \beta_{0,h} + \sum_{k=1}^{22} w_{k,h} \sigma_{t-k+1}$ with effective lag weights $w_{1,h} = \beta_{1,h} + \beta_{2,h}/5 + \beta_{3,h}/22$, $w_{2..5,h} = \beta_{2,h}/5 + \beta_{3,h}/22$, and $w_{6..22,h} = \beta_{3,h}/22$, independent compilation error increases expected forecast MSE by $(1 + \kappa_{\mathrm{in}})\sigma_{\mathrm{comp}}^2$, where $1$ comes from the target session and $\kappa_{\mathrm{in}} = \frac{1}{H}\sum_{h=1}^{H}\sum_{k=1}^{22} w_{k,h}^2 = 0.131$ comes from the multi-lag HAR-IV input filter (yielding a total multiplier of $1 + \kappa_{\mathrm{in}} = 1.131$, compared with $1 + 1 = 2.0$ for single-lag persistence). Using the $\alpha = 0.60$ grid compilation errors $\sigma_{\mathrm{comp}}$ from the $2019\text{--}2021$ 12-cell surface experiment in Table~\ref{tab:cross} as illustrative proxy noise levels (noting that the $2024\text{--}2025$ test index is a daily full-chain average rather than a re-subsampled 12-cell grid), target-aligned compilation ($\sigma_{\mathrm{comp}} = 11.7\text{ bps}$ under LW-PPS on LW or $31.3\text{ bps}$ under Plug-in Equal-Cell on EW) yields an analytical HAR-IV RMSE of $\sqrt{1{,}254^2 + 1.131\,\sigma_{\mathrm{comp}}^2} \approx 1{,}254.1\text{ bps}$ and $1{,}254.4\text{ bps}$ ($1{,}254.1\text{ bps}$ and $1{,}254.8\text{ bps}$ under the conservative persistence multiplier $2\sigma_{\mathrm{comp}}^2$), whereas mismatched LW-PPS compilation on the EW target ($\sigma_{\mathrm{comp}} = 186.3\text{ bps}$) raises the analytical RMSE to $1{,}269.6\text{ bps}$ ($1{,}281.4\text{ bps}$ under $2\sigma_{\mathrm{comp}}^2$).

\section{Multi-Horizon Risk Forecasting Models and Empirical Benchmarks}
\label{sec:models}

\subsection{Autoregressive Encoder-Decoder LSTM and Econometric Baselines}

Given an input window of $T_{\mathrm{in}} = 21$ hourly bars with standardized feature vectors $\mathbf{x}_{t - T_{\mathrm{in}} + 1 : t} \in \mathbb{R}^{T_{\mathrm{in}} \times D}$, our forecasting task is to jointly predict the 14-hour-ahead split-adjusted log-return path $r_{t,h} = \log(P_{t+h} / P_t)$ (recoverable in dollars as $\hat{P}_{t+h} = P_t \exp(\hat{r}_{t,h})$) and the forward EOD implied volatility path $\sigma_{t+h}$ across horizons $h = 1, \dots, H$ ($H = 14$ hourly bars, spanning two trading sessions). Specifically, at anchor hour $t$ on trading session $d(t)$, the latest known implied volatility in the input feature vector is the prior session's EOD implied volatility $\sigma_{t} = \mathrm{IV}_{d(t)-1}^{\mathrm{EOD}}$ (lagged by one full trading session to prevent intraday look-ahead), whereas the forward target at step $h \in \{1, \dots, 14\}$ is the EOD implied volatility $\sigma_{t+h} = \mathrm{IV}_{d(t+h)}^{\mathrm{EOD}}$ of the trading session $d(t+h)$ on which hour $t+h$ falls. Across all $N_{\mathrm{test}} \times H = 1{,}963 \times 14 = 27{,}482$ test forecast pairs, $0$ pairs ($0.00\%$) target the already-known $\mathrm{IV}_{d(t)-1}^{\mathrm{EOD}}$; $5{,}883$ pairs ($21.41\%$, where $t+h$ remains on session $d(t)$) forecast the current session's upcoming 16:00 ET closing $\mathrm{IV}_{d(t)}^{\mathrm{EOD}}$ (one session ahead of $\sigma_t$), and $21{,}599$ pairs ($78.59\%$, where $t+h$ crosses into session $d(t)+1$ or $d(t)+2$, including all $14$ horizons at the daily 15:00 ET session-close anchor bar) forecast strictly future sessions' EOD implied volatility ($\mathrm{IV}_{d(t)+1}^{\mathrm{EOD}}$ and $\mathrm{IV}_{d(t)+2}^{\mathrm{EOD}}$, two to three sessions ahead of $\sigma_t$).

Our primary neural architecture is a sequence-to-sequence Autoregressive Encoder-Decoder LSTM \citep{hochreiter1997long, sutskever2014sequence}. A 2-layer encoder LSTM (hidden dimension $d_h = 32$, dropout $0.15$) processes $\mathbf{x}_{t - T_{\mathrm{in}} + 1 : t}$ to produce a terminal context state $(\mathbf{h}_E, \mathbf{c}_E)$. A dedicated decoder LSTM initialized at $(\mathbf{h}_E, \mathbf{c}_E)$ unrolls autoregressively across $h = 1, \dots, H$: at each horizon step $h$, a learned horizon embedding $\mathbf{e}_h \in \mathbb{R}^{d_h}$ is combined with the previous step's projected prediction $(\hat{r}_{t,h-1}, \Delta\hat{\sigma}_{t,h-1})$ to update the decoder cell and emit standardized return and IV-change predictions, from which level forecasts are recovered as $\hat{\sigma}_{t+h} = \sigma_t + \Delta\hat{\sigma}_{t,h}$.

On any longitudinal subsample $\mathcal{S} \subset \mathcal{U}_{\mathrm{train}}$, the network parameters are trained using the AdamW optimizer \citep{loshchilov2019decoupled} (learning rate $10^{-3}$, weight decay $10^{-4}$, batch size $256$, gradient norm clipping at $1.0$, early stopping with patience $6$ on validation loss) under the tempered inverse-probability weighted multi-task Huber and quadratic IV-change loss:
\begin{equation}
    \mathcal{L}_{\mathrm{total}}(\theta) = \frac{1}{|\mathcal{S}|} \sum_{t \in \mathcal{S}} \tilde{d}_t \left[ \frac{1}{H}\sum_{h=1}^{H} \ell_{\delta}\!\left(\frac{\hat{r}_{t,h} - r_{t,h}}{s_{r,h}}\right) + \frac{\eta}{H}\sum_{h=1}^{H} \left(\frac{\hat{\sigma}_{t+h} - \sigma_{t+h}}{s_{\sigma,h}}\right)^2 \right],
    \label{eq:loss}
\end{equation}
where $\tilde{d}_t = |\mathcal{S}| \, \pi_t^{-\gamma_{\mathrm{HT}}} / \sum_{j \in \mathcal{S}} \pi_j^{-\gamma_{\mathrm{HT}}}$ is the sample-normalized tempered design weight ($\gamma_{\mathrm{HT}} = 0.55$ in the baseline and $\gamma_{\mathrm{HT}} \in \{0.00, 0.55, 1.00\}$ in our weight-tempering sensitivity ablation), $\ell_{\delta}(\cdot)$ is the Huber loss ($\delta = 1.0$), $s_{r,h}$ and $s_{\sigma,h}$ are training-split target standard deviations at horizon $h$, and $\eta = 0.25$ balances the return and implied-volatility objectives. Note that because $\sigma_{t+h}$ represents a cross-sectionally aggregated daily chain IV index stepped across forecast time $t+h$ rather than a fixed-date maturity slice $\tau \mapsto \sigma_{\mathrm{imp}}(m, \tau; t)$ at fixed origin $t$, static calendar-spread total-variance inequalities \citep{gatheral2004parsimonious} do not constrain the temporal trajectory $h \mapsto \sigma_{t+h}$ and are deliberately omitted from Equation~\eqref{eq:loss}.

We benchmark the Encoder-Decoder LSTM (evaluated across 10 independent random seeds per sampling design, reporting both seed mean $\pm$ SD and equal-weight seed-ensemble predictions) against four reference models. The first is the No-Change Random Walk and IV Persistence baseline, which sets $\hat{P}_{t+h} = P_t$ ($\hat{r}_{t,h} = 0$) for prices and $\hat{\sigma}_{t+h} = \sigma_t$ for implied volatility. The second is the Historical Drift baseline $\hat{r}_{t,h} = \bar{r}_{\mathrm{train},h}$, where $\bar{r}_{\mathrm{train},h} > 0$ is the training-split mean $h$-hour log-return. The third is Linear Ridge Regression using multi-output $L_2$-penalized regression ($\lambda \in \{1, 10, \dots, 10^5\}$ selected on validation MSE) fit on lookback summaries using either all features ($D=31$, $\lambda = 10^3$) or price and technical features ($D=16$, $\lambda = 10^5$). The fourth is our Heterogeneous Autoregressive Implied Volatility specification (HAR-IV), which adapts the daily, weekly, and monthly lag structure of the HAR-RV realized volatility model of \citet{corsi2009simple} to our daily implied-volatility index by fitting an ordinary least squares regression on the training split at each horizon $h = 1, \dots, 14$ predicting forward EOD implied volatility $\sigma_{t+h}$ from the current (1-session-lagged) EOD implied volatility $\sigma_t$, its 5-session trailing mean $\bar{\sigma}_{t}^{(5\mathrm{d})}$, and its 22-session trailing mean $\bar{\sigma}_{t}^{(22\mathrm{d})}$:
\begin{equation}
    \hat{\sigma}_{t+h}^{\mathrm{HAR}} = \beta_{0,h} + \beta_{1,h} \sigma_t + \beta_{2,h} \bar{\sigma}_{t}^{(5\mathrm{d})} + \beta_{3,h} \bar{\sigma}_{t}^{(22\mathrm{d})}.
    \label{eq:har_iv}
\end{equation}

\subsection{Compute-Efficient Subsampling vs.\ Full Census: Paired TOST Equivalence and Price Benchmark}

\begin{table}[!htbp]
\centering
\caption{Paired 10-Seed Comparison and TOST Equivalence of Probability Subsampling Designs vs.\ Full Census for the Encoder-Decoder LSTM on Tier 5a ($D = 31$, All Features + OptionVAE Latents; $N_{\mathrm{test}} = 1{,}963$ Windows).}
\label{tab:subsampling}
\resizebox{\textwidth}{!}{\IfFileExists{tables/tab_subsampling.tex}{\input{tables/tab_subsampling.tex}}{\input{tab_subsampling.tex}}}
\vspace{2pt}
\begin{minipage}{0.98\textwidth}
\footnotesize \textit{Notes:} Reports the paired percentage change $\Delta\%$ in multi-horizon Price RMSE (versus full-census mean $\$4.156$) and Implied Volatility RMSE (versus full-census mean $1{,}377\text{ bps}$) across $S=10$ matched random seeds on the comprehensive 31-feature Tier 5a specification ($D = 31$, all features plus OptionVAE latents; Section~\ref{sec:iv_ablation} Table~\ref{tab:ablation} subsequently shows that pruning the 8 noisy activity aggregates in Tier 2 [$D=20$] and Tier 5b [$D=23$] improves absolute IV skill while preserving the same subsampling behavior), $90\%$ Two One-Sided Tests (TOST) confidence intervals $[L_{90}, U_{90}]$, two-sided paired $t$-test $p$-values ($p_{\mathrm{diff}}$ for $H_0\text{: }\mu_{\Delta} = 0$), formal TOST equivalence $p$-values ($p_{\mathrm{TOST}}$ for $H_0\text{: }|\mu_{\Delta}| \ge \Delta_{\mathrm{eq}}$ vs.\ $H_1\text{: }|\mu_{\Delta}| < \Delta_{\mathrm{eq}}$ under chosen economically motivated equivalence margins $\Delta_{\mathrm{eq}}^{\mathrm{price}} = \pm 0.50\%$ [$\pm \$0.021$] and $\Delta_{\mathrm{eq}}^{\mathrm{IV}} = \pm 2.00\%$ [$\pm 27.5\text{ bps}$]; \citealp{schuirmann1987comparison, lakens2017equivalence}), and wall-clock training time reduction $\Delta\%$. At $\alpha = 0.60$, all three probability designs, including SRS, satisfy formal statistical equivalence relative to the full census ($p_{\mathrm{TOST}} \le 0.002$ for price and $p_{\mathrm{TOST}} \le 0.020$ for IV) while cutting training time by $21\%\text{--}22\%$; direct paired differences of Liquidity PPS and Neyman against SRS at $\alpha = 0.60$ are $+0.02\%$ ($p=0.893$) and $+0.01\%$ ($p=0.963$) on price RMSE, and $+0.51\%$ ($p=0.608$) and $+0.09\%$ ($p=0.937$) on IV RMSE.
\end{minipage}
\end{table}

\begin{figure}[!htbp]
    \centering
    \includegraphics[width=0.92\textwidth]{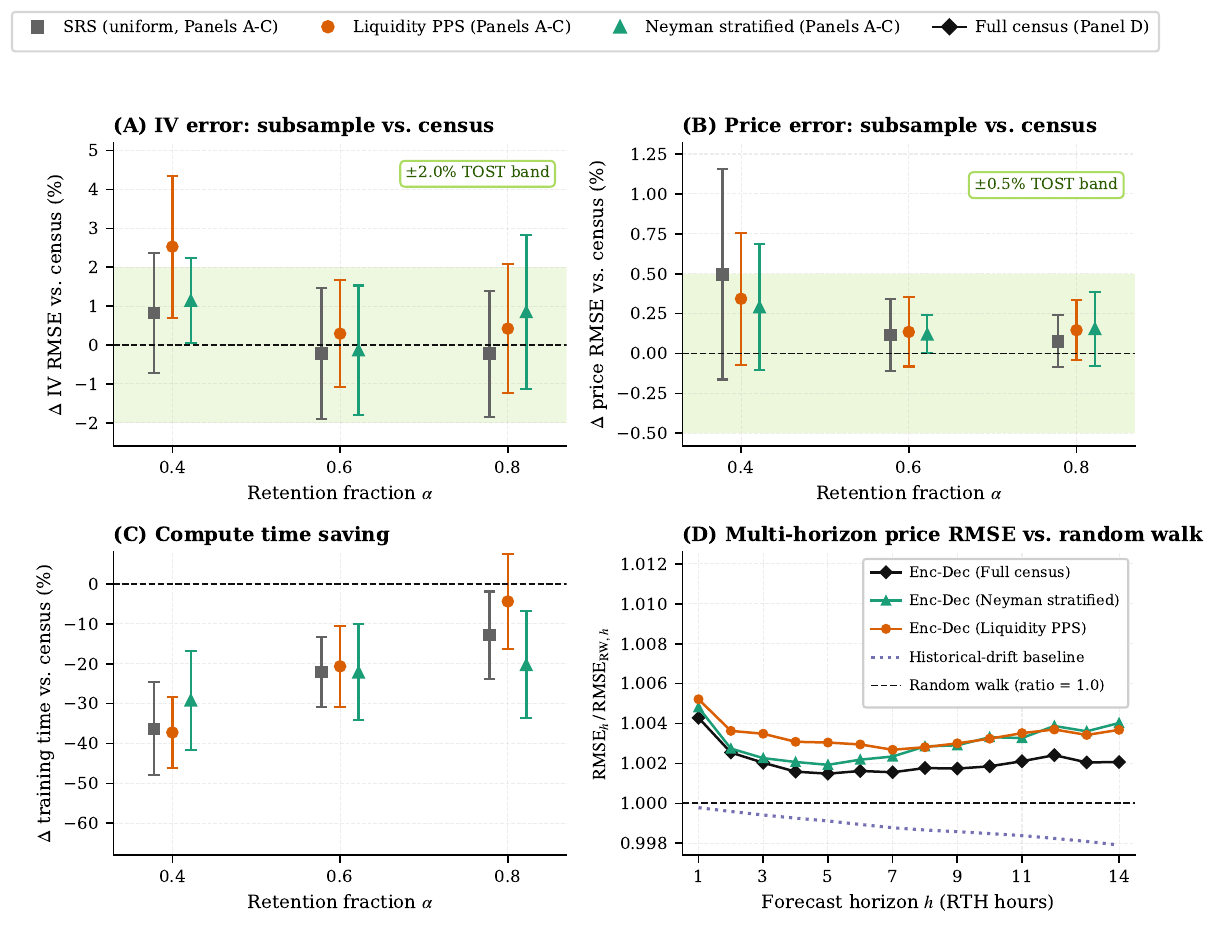}
    \caption{Longitudinal Subsampling Efficiency, TOST Equivalence Bands, and Random-Walk Price Benchmark on the Encoder-Decoder LSTM (Tier 5a, $D = 31$, 10 Paired Seeds). Panel A: Paired percentage change in out-of-sample IV forecast RMSE relative to the unweighted full census ($95\%$ CI error bars across 10 seeds and shaded $\pm 2.0\%$ TOST equivalence band). Panel B: Paired percentage change in Price RMSE relative to the full census with shaded $\pm 0.5\%$ TOST equivalence band. Panel C: Wall-clock training time reduction (\%) across $\alpha \in \{0.4, 0.6, 0.8\}$. Panel D: Ratio of model price RMSE to the no-change Random Walk RMSE across forecast horizons $h = 1, \dots, 14$ RTH hours.}
    \label{fig:training_subsampling}
\end{figure}

Because failing to reject the null of zero difference ($p_{\mathrm{diff}} > 0.05$) does not by itself establish statistical equivalence, Table~\ref{tab:subsampling} and Figure~\ref{fig:training_subsampling} (Panels A--C) evaluate longitudinal survey subsampling on the Encoder-Decoder LSTM across 10 paired random seeds on the comprehensive 31-feature Tier 5a specification ($D = 31$, combining all 28 spot/options features with the 3 OptionVAE latents; Section~\ref{sec:iv_ablation} Table~\ref{tab:ablation} then evaluates pruned 20- and 23-feature representations) using both two-sided difference tests ($p_{\mathrm{diff}}$) and formal Two One-Sided Tests (TOST; \citealp{schuirmann1987comparison, lakens2017equivalence}) under chosen practical equivalence bands of $\Delta_{\mathrm{eq}}^{\mathrm{price}} = \pm 0.50\%$ ($\pm \$0.021$ on a $\$4.156$ RMSE baseline, comparable to a two-cent equity tick increment) and $\Delta_{\mathrm{eq}}^{\mathrm{IV}} = \pm 2.00\%$ ($\pm 27.5\text{ bps}$ on a $1{,}377\text{ bps}$ baseline, less than half of a typical single-contract OTM option bid-ask implied volatility spread). Across retention levels $\alpha \in \{0.40, 0.60, 0.80\}$, subsampling cuts wall-clock training time by $29\%$ to $37\%$ at $\alpha = 0.40$, by $21\%$ to $22\%$ at $\alpha = 0.60$, and by $4\%$ to $20\%$ at $\alpha = 0.80$ (Panel C of Figure~\ref{fig:training_subsampling}). For multi-horizon price forecasting at $\alpha \in \{0.60, 0.80\}$, all three sampling designs keep mean price RMSE within $[+0.08\%, +0.15\%]$ of the full census and pass formal TOST equivalence at the $1\%$ level ($p_{\mathrm{TOST}} \in [0.00004, 0.0041]$, with all $90\%$ TOST confidence intervals strictly inside $[-0.50\%, +0.50\%]$; Panel B). Note that for Neyman stratification at $\alpha = 0.60$, the two-sided test yields $p_{\mathrm{diff}} = 0.048$ solely because cross-seed variance is tiny ($90\%$ CI $[+0.02\%, +0.22\%]$ around $+0.12\%$, or $+\$0.005$), while TOST confirms practical equivalence with $p_{\mathrm{TOST}} < 0.001$.

For implied volatility forecasting on Tier 5a ($D = 31$), once the 2024--2025 option tape is harmonized via full-chain OPRA.PILLAR snapshots, unweighted full-census training achieves a 10-seed mean IV RMSE of $1{,}377\text{ bps}$ ($1{,}346\text{ bps}$ ensemble, $+8.9\%$ MSE skill). At $\alpha = 0.60$, tempered inverse-probability weighted subsampling ($\gamma_{\mathrm{HT}} = 0.55$) passes formal TOST equivalence relative to the full census within the $\pm 2.00\%$ margin across every design, including SRS: paired $\Delta\mathrm{IV\ RMSE}$ versus the full census is $-0.13\%$ ($90\%$ TOST CI $[-1.48\%, +1.21\%]$, $p_{\mathrm{diff}} = 0.860$, $p_{\mathrm{TOST}} = 0.016$) under Neyman stratification, $+0.29\%$ ($90\%$ CI $[-0.82\%, +1.40\%]$, $p_{\mathrm{diff}} = 0.642$, $p_{\mathrm{TOST}} = 0.010$) under Liquidity PPS, and $-0.22\%$ ($90\%$ CI $[-1.58\%, +1.14\%]$, $p_{\mathrm{diff}} = 0.775$, $p_{\mathrm{TOST}} = 0.020$) under SRS. Moreover, the 10-seed ensemble forecast under Neyman stratification ($\alpha = 0.60$, Tier 5a) achieves $1{,}339\text{ bps}$ ($+9.9\%$ MSE skill), matching the full-census Tier 5a ensemble ($1{,}346\text{ bps}$, $+8.9\%$ MSE skill; a $-0.5\%$ difference well inside the $\pm 2.00\%$ equivalence margin) while using $40\%$ fewer training windows. At the same time, we separate this subsampling-versus-census equivalence from the comparison between informative longitudinal designs and SRS at the same sample budget: while Liquidity PPS and Neyman stratification retain substantially more top-decile realized-volatility windows than SRS ($81.2\%$ and $73.7\%$ vs.\ $59.9\%$ at $\alpha = 0.60$; Table~\ref{tab:sampling}), direct paired 10-seed comparisons against SRS at $\alpha = 0.60$ yield differences of $+0.02\%$ ($p_{\mathrm{diff}} = 0.893$) for PPS and $+0.01\%$ ($p_{\mathrm{diff}} = 0.963$) for Neyman on price RMSE, and $+0.51\%$ ($p_{\mathrm{diff}} = 0.608$) for PPS and $+0.09\%$ ($p_{\mathrm{diff}} = 0.937$) for Neyman on IV RMSE. Thus, all three probability designs preserve full-census accuracy at $60\%$ retention, whereas incremental out-of-sample forecasting gains of informative longitudinal selection over SRS remain unestablished. At aggressive $40\%$ retention ($\alpha = 0.40$), IV RMSE degrades relative to the census by $+1.14\%$ ($p_{\mathrm{diff}} = 0.042, p_{\mathrm{TOST}} = 0.054$) under Neyman and $+2.53\%$ ($p_{\mathrm{diff}} = 0.012, p_{\mathrm{TOST}} = 0.736$) under PPS (Panel A of Figure~\ref{fig:training_subsampling}), identifying $\alpha = 0.60$ as the practical retention threshold.

\begin{table}[!htbp]
\centering
\caption{Multi-Horizon Equity Price Forecasting Benchmark on the Leave-Out Test Set ($N_{\mathrm{test}} = 1{,}963$ Windows Across $281$ Sessions, $H = 14$ Hours Ahead).}
\label{tab:price}
{\small\setlength{\tabcolsep}{3.5pt}\IfFileExists{tables/tab_price.tex}{\input{tables/tab_price.tex}}{\input{tab_price.tex}}}
\vspace{2pt}
\begin{minipage}{0.98\textwidth}
\footnotesize \textit{Notes:} Reports pooled price RMSE (\$), horizon-specific RMSE at $h \in \{1, 7, 14\}$ hours (\$), out-of-sample Forecast Skill Score ($\mathrm{FSS} = 1 - \mathrm{MSE}/\mathrm{MSE}_{\mathrm{RW}}$, \%), \citet{diebold1995comparing} $t$-statistic vs.\ the random walk at $h=7$ computed with \citet{flegal2010batch, liu2022lugsail, flegal2025mcmcse} \texttt{mcmcse} lugsail batch-means time-average standard errors (where $t_{\mathrm{DM}} > 0$ indicates higher squared error, i.e., worse accuracy, than the random walk under $d_t = e_{\mathrm{model},t}^2 - e_{\mathrm{RW},t}^2$), and 1-day ($h=7$) directional accuracy ($\mathrm{DA}_{7\mathrm{h}}$, \%). Panel B reports mean $\pm$ SD across $S=10$ random seeds on the 31-feature Tier 5a specification ($D=31$, all features plus OptionVAE latents). Across all neural configurations, $|t_{\mathrm{DM}}| \le 0.50$ ($|t_{\mathrm{DM}}| \le 1.27$ including dense Linear Ridge), indicating that short-horizon prices are not forecastable beyond a random walk with drift among tested models.
\end{minipage}
\end{table}

Table~\ref{tab:price} and Panel D of Figure~\ref{fig:training_subsampling} confirm a central empirical regularity of liquid equity markets: across the tested linear and recurrent architectures, 1-to-14 hour split-adjusted NVDA price levels are not forecastable out-of-sample beyond a random walk with positive drift. On the leave-out test set ($N_{\mathrm{test}} = 1{,}963$ windows), the no-change Random Walk achieves a pooled RMSE of $\$4.148$ ($\mathrm{RMSE}_{1\mathrm{h}} = \$1.516, \mathrm{RMSE}_{7\mathrm{h}} = \$4.013, \mathrm{RMSE}_{14\mathrm{h}} = \$5.636$), and the Historical Drift baseline achieves $\$4.141$ ($\mathrm{FSS} = +0.30\%$, \citealp{diebold1995comparing} $t_{\mathrm{DM}} = -0.47$ at $h=7$ with \citealp{flegal2010batch, liu2022lugsail, flegal2025mcmcse} \texttt{mcmcse} batch-means standard errors). Across all sampling designs, the Encoder-Decoder LSTM (Tier 5a, $D=31$) achieves a pooled price RMSE of $\$4.156\text{--}\$4.161$ ($\mathrm{FSS} \in [-0.66\%, -0.39\%]$, $|t_{\mathrm{DM}}| \le 0.50$), statistically indistinguishable from the random walk while outperforming dense Linear Ridge regression on all 31 features ($\$4.207$, $\mathrm{FSS} = -2.90\%$, $t_{\mathrm{DM}} = +1.27$). Because NVDA appreciated over the 2024--2025 test period, a passive always-up classifier attains $55.17\%$ 1-day ($h=7$) directional accuracy, matched closely by the Encoder-Decoder LSTM ($54.7\%\text{--}55.0\%$).

\subsection{Implied Volatility Risk Forecasting, Session-Clustered Inference, and Representation Ablation}
\label{sec:iv_ablation}

\begin{figure}[!htbp]
    \centering
    \includegraphics[width=0.92\textwidth]{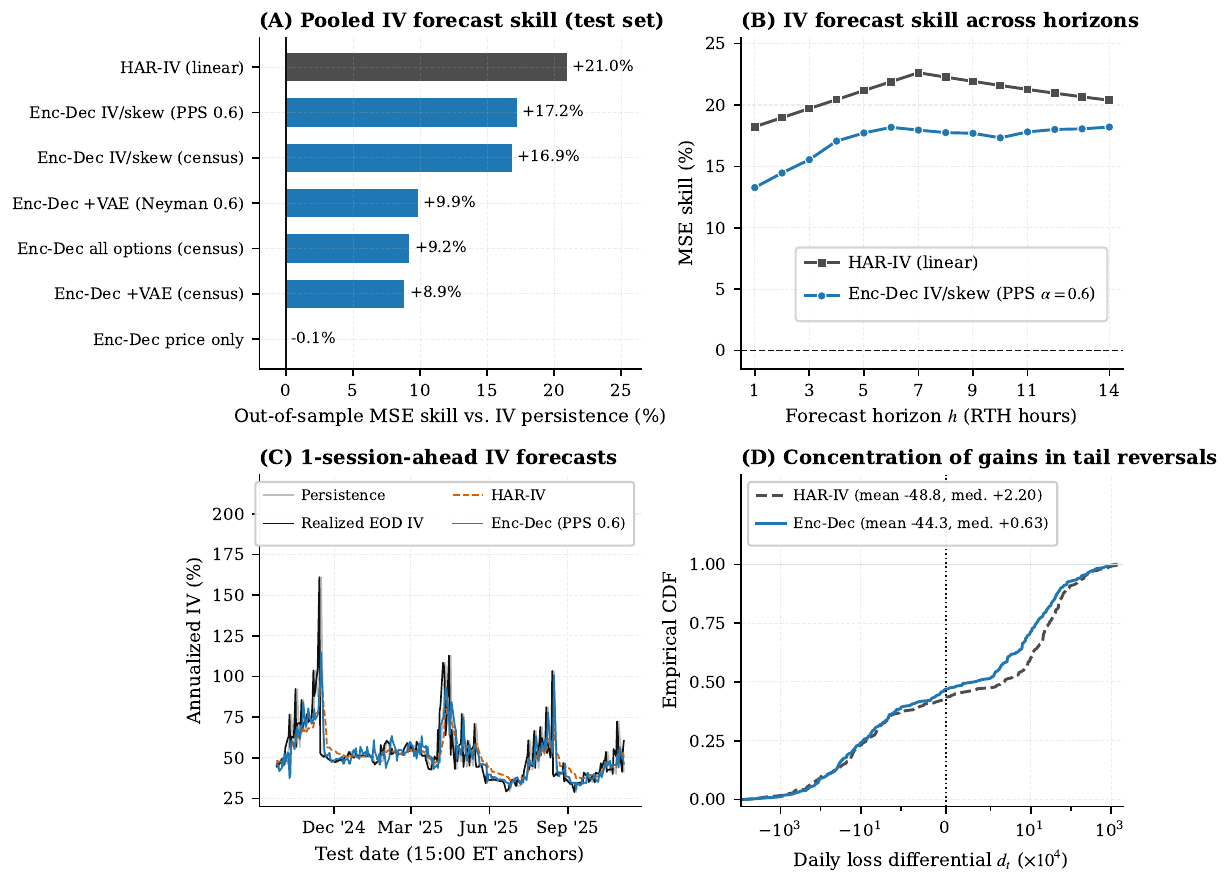}
    \caption{Out-of-Sample Implied Volatility Forecasting Skill and Loss-Differential Diagnostics ($2024\text{--}2025$ Leave-Out Test Set, $G = 281$ Sessions). Panel A: Pooled out-of-sample IV MSE skill (\%) relative to persistence across linear HAR-IV and Encoder-Decoder LSTM feature and sampling configurations. Panel B: Horizon-by-horizon IV MSE skill (\%) across $h = 1, \dots, 14$ hours for HAR-IV and Encoder-Decoder LSTM (Tier 2 IV/skew/VRP features, Liquidity PPS $\alpha=0.60$). Panel C: One-session-ahead (15:00 ET anchor) realized vs.\ predicted EOD implied volatility trajectories across the test window. Panel D: Empirical cumulative distribution function (ECDF) of daily squared-error differentials $d_t = e_{\mathrm{model},t}^2 - e_{\mathrm{persist},t}^2$ ($\times 10^4$, symmetric log scale), showing positive sample median differential ($+2.20$ and $+0.63$) alongside strongly negative mean differential ($-48.8$ and $-44.3$) driven by post-spike mean reversion.}
    \label{fig:iv_forecasting}
\end{figure}

In contrast to short-horizon equity price changes, option-implied volatility is a mean-reverting process whose spikes following earnings announcements and macro sell-offs decay over subsequent sessions \citep{andersen2000benchmarking, goncalves2006predictable, corsi2009simple, bernales2014can}. Table~\ref{tab:iv}, Table~\ref{tab:ablation}, and Figure~\ref{fig:iv_forecasting} evaluate multi-horizon forecasting of forward EOD implied volatility across the $N_{\mathrm{test}} = 1{,}963$ test windows ($G = 281$ trading sessions). Because EOD implied volatility is recorded once per session and repeated across the 7 intraday anchor hours of each session, testing all $1{,}963$ overlapping hourly windows as independent observations would understate standard errors. We therefore pair pooled multi-horizon RMSE and MSE skill with strictly future-session target skill ($d(t+h) > d(t)$, $21{,}599$ pairs), pure daily 1- and 2-session-ahead HAR-IV regressions on unique trading sessions (Panel C of Table~\ref{tab:iv}), and three session-level inference procedures: (i) a session-clustered pooled \citet{diebold1995comparing} test ($t_{\mathrm{sess}}$), which averages the 14-horizon squared-error differential $d_{t} = \frac{1}{H}\sum_{h=1}^H (e_{\mathrm{model},t,h}^2 - e_{\mathrm{bench},t,h}^2)$ within each of the $G = 281$ test sessions to form a single session-level loss differential $\bar{d}_g$ and estimates the time-average standard error $\mathrm{SE}_{\mathrm{BM}}(\bar{d})$ across sessions using the data-driven lugsail batch-means estimator of \citet{flegal2010batch}, \citet{liu2022lugsail}, and \citet{flegal2025mcmcse} (\texttt{mcmcse} v1.5-1), evaluated against both IV persistence and linear HAR-IV (overlapping batch-means and Bartlett spectral-variance estimators in \texttt{mcmcse} give closely matching results: $t_{\mathrm{sess}} = -1.04$ [$p=0.30$] for HAR-IV, $-1.03$ [$p=0.30$] for Tier 2 PPS, and $-1.25$ [$p=0.21$] for Tier 2 Census); (ii) an outlier-trimmed session-clustered DM test ($t_{\mathrm{trim}}$) removing the two extreme 2025 earnings/tariff jump sessions; and (iii) a non-overlapping daily 15:00 ET anchor evaluation ($281$ daily anchors) reporting the daily \texttt{mcmcse} batch-means DM test ($t_{15\mathrm{h}}$), the percentage of sessions favoring persistence ($d_g > 0$) with its 5-session moving-block bootstrap \citep{kunsch1989jackknife} $95\%$ confidence interval ($B = 2{,}000$), both two-sided and one-sided ($H_1\text{: }\text{model} < \text{persist}$) \citet{wilcoxon1945individual} signed-rank $p$-values, and the share of cumulative net squared-error gains concentrated in the top 5 volatility-reversal sessions.

\begin{table}[!htbp]
\centering
\caption{Out-of-Sample Implied Volatility Forecasting Benchmark, Session-Clustered Inference, and Tail Concentration ($N_{\mathrm{test}} = 1{,}963$ Windows Across $G = 281$ Test Sessions).}
\label{tab:iv}
\resizebox{\textwidth}{!}{\IfFileExists{tables/tab_iv.tex}{\input{tables/tab_iv.tex}}{\input{tab_iv.tex}}}
\vspace{2pt}
\begin{minipage}{0.98\textwidth}
\footnotesize \textit{Notes:} Reports pooled multi-horizon IV RMSE (in basis points of annualized IV, $100\text{ bps} = 1\%$ IV), pooled out-of-sample MSE skill ($1 - \mathrm{MSE}/\mathrm{MSE}_{\mathrm{persist}}$, \%) across all $27{,}482$ test pairs ($0\%$ of which target the known lagged $\mathrm{IV}_{d-1}$), strictly future-session MSE skill across the $21{,}599$ pairs targeting $\mathrm{IV}_{d+1}$ or $\mathrm{IV}_{d+2}$ ($d(t+h) > d(t)$), session-clustered \citet{diebold1995comparing} $t_{\mathrm{sess}}$ ($p$-val) vs.\ persistence across all $G = 281$ test sessions computed using \citet{flegal2010batch, liu2022lugsail, flegal2025mcmcse} \texttt{mcmcse} batch-means time-average standard errors (overlapping batch-means and Bartlett spectral-variance estimators in \texttt{mcmcse} yield closely matching statistics: $t_{\mathrm{sess}} = -1.04$ [$p=0.30$] for HAR-IV, $-1.03$ [$p=0.30$] for Tier 2 PPS, and $-1.25$ [$p=0.21$] for Tier 2 Census), outlier-trimmed session DM $t_{\mathrm{trim}}$ ($p$-val) excluding the two extreme 2025 jump sessions, direct session-clustered DM $t_{\mathrm{sess}}$ ($p$-val) vs.\ linear HAR-IV (positive values indicate higher MSE than HAR-IV), the sample median daily 15:00 ET squared-error differential $d_t = e_{\mathrm{model},t}^2 - e_{\mathrm{persist},t}^2$ ($\times 10^4$, or $\text{bps}^2$; daily 15:00 ET \texttt{mcmcse} DM $t_{15\mathrm{h}}$ is $-1.01$ [$p=0.31$] for HAR-IV, $-1.13$ [$p=0.26$] for Tier 2 PPS, and $-1.30$ [$p=0.19$] for Tier 2 Census), and the percentage of 15:00 ET sessions with $d_t > 0$ (where persistence has lower squared error than the model) with $95\%$ 5-session moving-block bootstrap CIs ($B=2{,}000$; on session-mean differentials $\bar{d}_g > 0$, persistence wins on $58.7\%$ $[51.2\%, 66.5\%]$ of sessions for HAR-IV and $60.1\%$ $[54.1\%, 66.5\%]$ for Tier 2 PPS, while two-sided / one-sided \citealp{wilcoxon1945individual} signed-rank $p$-values on 15:00 ET $d_t$ are $0.008 / 1.00$ for HAR-IV, $0.35 / 0.82$ for Tier 2 Census, $0.27 / 0.87$ for Tier 2 PPS, $0.52 / 0.74$ for Tier 5a Census, and $0.29 / 0.86$ for Tier 5a Neyman). The top 5 post-spike volatility-reversal sessions contribute $157\%$ (HAR-IV), $144\%$ (Tier 2 PPS), and $119\%$ (Tier 2 Census) of net cumulative squared-error reduction. Panel C reports pure daily session-ahead HAR-IV regressions fit on unique daily closing sessions ($G = 281$ for $\mathrm{IV}_{d+1}$, persistence RMSE $1{,}194\text{ bps}$; $G = 280$ for $\mathrm{IV}_{d+2}$, persistence RMSE $1{,}433\text{ bps}$).
\end{minipage}
\end{table}

\begin{table}[!htbp]
\centering
\caption{Feature Representation (Augmentation vs.\ Replacement) and Weight-Tempering Ablation on the Encoder-Decoder LSTM (5-Seed Mean $\pm$ SD).}
\label{tab:ablation}
\resizebox{\textwidth}{!}{\IfFileExists{tables/tab_ablation.tex}{\input{tables/tab_ablation.tex}}{\input{tab_ablation.tex}}}
\vspace{2pt}
\begin{minipage}{0.98\textwidth}
\footnotesize \textit{Notes:} Panel A compares out-of-sample Price Forecast Skill Score ($\mathrm{FSS}$ vs.\ random walk, \%) and Implied Volatility MSE Skill (vs.\ IV persistence, \%) across nested feature tiers under unweighted Full Census ($\alpha=1.00$) and tempered inverse-probability weighted Liquidity PPS ($\alpha=0.60, \gamma_{\mathrm{HT}}=0.55$), contrasting 31-feature latent \emph{augmentation} (Tiers 4a and 5a, which append 3 PCA or OptionVAE latents to all 28 features of Tier 3) against 23-feature latent \emph{replacement} (Tiers 4b and 5b, which replace the 8 noisy option activity aggregates with the 3 PCA or OptionVAE latents). Panel B reports the sensitivity at $\alpha = 0.60$ to the inverse-probability weight-tempering exponent $\gamma_{\mathrm{HT}} \in \{0.00, 0.55, 1.00\}$ (where $\gamma_{\mathrm{HT}} = 1.00$ after sample-mean normalization is self-normalized H\'{a}jek ratio IPW) under Neyman stratification (left columns, Tier 5a) and Liquidity PPS (right columns, reporting Tier 5a and, in parentheses, Tier 5b replacement). Entries report mean $\pm$ SD across 5 matched random seeds.
\end{minipage}
\end{table}

Table~\ref{tab:iv}, Table~\ref{tab:ablation}, and Figure~\ref{fig:iv_forecasting} establish three principal findings on implied volatility risk predictability and representation design. First, relative to the IV persistence baseline ($\mathrm{RMSE} = 1{,}410\text{ bps}$), linear HAR-IV achieves a pooled RMSE of $1{,}254\text{ bps}$, corresponding to a $+21.0\%$ out-of-sample MSE reduction ($+18.3\%$ at $h=1\text{h}$, $+22.7\%$ at $h=7\text{h}$, and $+20.4\%$ at $h=14\text{h}$; Panel B of Figure~\ref{fig:iv_forecasting}). On the Encoder-Decoder LSTM, the parsimonious Tier 2 specification (Price + lagged IV, Skew, $\Delta\mathrm{IV}_{1\mathrm{d}}$, and VRP proxy, $D=20$) trained under Liquidity PPS ($\alpha = 0.60$) achieves an ensemble RMSE of $1{,}283\text{ bps}$ ($+17.2\%$ MSE skill), and full-census Tier 2 training achieves $1{,}286\text{ bps}$ ($+16.9\%$ ensemble skill). Importantly, comparing the neural sequence models directly against linear HAR-IV shows that the Encoder-Decoder LSTM does not outperform the simpler linear HAR-IV benchmark: on Tier 2, the LSTM ensemble trails HAR-IV by $29\text{--}32\text{ bps}$ ($1{,}283\text{--}1{,}286\text{ bps}$ vs.\ $1{,}254\text{ bps}$), with direct session-clustered Diebold-Mariano statistics against HAR-IV of $t_{\mathrm{sess}} = +0.51$ ($p = 0.61$) for Tier 2 PPS and $t_{\mathrm{sess}} = +0.46$ ($p = 0.64$) for Tier 2 Census ($t_{\mathrm{sess}} = +1.42, p = 0.16$ for Tier 5a Neyman, $+1.35, p = 0.18$ for Tier 3 Census, and $+1.48, p = 0.14$ for Tier 5a Census; Table~\ref{tab:iv}). Decomposing the $27{,}482$ test predictions by anchor hour and target session confirms that the descriptive MSE reductions over persistence are not an artifact of intraday target repetition: on the $21{,}599$ pairs ($78.6\%$) targeting strictly future sessions ($d(t+h) > d(t)$), MSE skill is $+22.0\%$ for HAR-IV, $+18.2\%$ for Tier 2 PPS, and $+17.9\%$ for Tier 2 census (excluding 15:00 ET anchors altogether yields $+20.7\%, +16.6\%,$ and $+16.5\%$, while 15:00 ET anchors alone yield $+22.5\%, +20.4\%,$ and $+19.2\%$). Likewise, Panel C of Table~\ref{tab:iv} shows that a pure daily session-ahead HAR-IV regression estimated on unique daily closes achieves $+17.4\%$ MSE skill at 1 session ahead ($\mathrm{IV}_{d+1}$, $1{,}085\text{ bps}$ vs.\ $1{,}194\text{ bps}$ persistence) and $+22.9\%$ MSE skill at 2 sessions ahead ($\mathrm{IV}_{d+2}$, $1{,}258\text{ bps}$ vs.\ $1{,}433\text{ bps}$).

Second, Panel A of Table~\ref{tab:ablation} clarifies how feature pruning and low-dimensional PCA and OptionVAE embeddings affect generalization on the harmonized 2024--2025 OPRA.PILLAR tape. Without lagged option-implied volatility (Tier 1, Price and Technicals only, $D=16$), the Encoder-Decoder LSTM cannot anchor the current volatility regime and achieves $-0.1\%\text{ to }-0.3\%$ MSE skill ($1{,}411\text{ bps}$). Adding the four core scale-invariant option metrics in Tier 2 ($D=20$) raises 5-seed mean IV skill to $+13.4\% \pm 3.7\%$ on the full census and $+11.0\% \pm 10.6\%$ under Liquidity PPS ($\alpha=0.60$). Adding all eight auxiliary option-activity aggregates in Tier 3 ($D=28$) degrades out-of-sample accuracy, reducing 5-seed mean IV skill to $+3.9\% \pm 5.1\%$ on the census ($1{,}344\text{ bps}$, $+9.2\%$ ensemble skill in Table~\ref{tab:iv}; paired $t = -3.15, p = 0.034$, Welch $p = 0.011$ vs.\ Tier 2) and $+1.8\% \pm 3.5\%$ under PPS. Appending 3 PCA or OptionVAE latents on top of all 28 features (augmentation Tiers 4a and 5a, $D=31$) only partially offsets this degradation ($+7.7\% \pm 1.1\%$ for PCA and $+3.0\% \pm 5.1\%$ for OptionVAE under PPS). By contrast, when the 8 noisy option-activity aggregates are removed and replaced by the 3 PCA or OptionVAE latents (replacement Tiers 4b and 5b, $D=23$), 5-seed mean IV skill is restored to $+13.0\% \pm 5.0\%$ (census) and $+12.3\% \pm 5.3\%$ (PPS; Welch $p = 0.008$ vs.\ Tier 3) for PCA (Tier 4b), and to $+11.8\% \pm 6.3\%$ (census) and $+13.6\% \pm 2.8\%$ (PPS; paired $p = 0.011$, Welch $p = 0.0004$ vs.\ Tier 3) for OptionVAE (Tier 5b). Importantly, relative to parsimonious Tier 2 ($D=20$, which simply drops the 8 noisy aggregates without adding latents), Tiers 4b and 5b are statistically indistinguishable in mean IV skill ($p_{\mathrm{pair}} \in [0.59, 0.86]$, $p_{\mathrm{Welch}} \in [0.61, 0.88]$). Thus, the primary accuracy gain comes from pruning the 8 noisy activity aggregates rather than latent extraction per se, while OptionVAE replacement ($D=23$) retains compressed surface-activity information and exhibits a lower descriptive 5-seed cross-seed standard deviation under Liquidity PPS ($2.8\%$ vs.\ $10.6\%$ for Tier 2).

Third, Panel B of Table~\ref{tab:ablation} evaluates the sensitivity of informative subsampling at $\alpha = 0.60$ to the inverse-probability weight-tempering exponent $\gamma_{\mathrm{HT}} \in \{0.00, 0.55, 1.00\}$. As shown by Equation~\eqref{eq:tempered_risk}, unweighted subsampling ($\gamma_{\mathrm{HT}} = 0.00$) shifts the design-expected training mean IV by $+398\text{ bps}$ under Liquidity PPS and $+253\text{ bps}$ under Neyman stratification ($+395\text{ bps}$ and $+250\text{ bps}$ in 10-seed realized sample averages). Empirically, increasing $\gamma_{\mathrm{HT}}$ from $0.00$ to $0.55$ (tempered IPW) and $1.00$ (self-normalized H\'{a}jek ratio IPW) produces a consistent directional improvement in 5-seed mean IV skill on both Tier 5a ($+0.1\% \pm 6.0\% \to +3.0\% \pm 5.1\% \to +5.5\% \pm 5.2\%$ under PPS, paired $t = 2.75, p = 0.051$, Welch $p = 0.167$; and $+1.8\% \to +3.2\% \to +3.2\%$ under Neyman, improving price FSS from $-1.44\%$ to $-0.54\%$ and $-0.35\%$) and Tier 5b replacement ($+12.5\% \pm 4.1\% \to +13.6\% \pm 2.8\% \to +14.0\% \pm 2.6\%$ under PPS, with descriptive cross-seed SD of $2.6\%\text{--}2.8\%$ vs.\ $4.1\%$, $p_{\mathrm{pair}} = 0.36\text{--}0.38$). While cross-seed variance at $S=5$ seeds precludes rejecting homogeneity across $\gamma_{\mathrm{HT}}$ at the $5\%$ level, tempered inverse-probability weighting ($\gamma_{\mathrm{HT}} = 0.55$) provides a principled theoretical safeguard against informative sampling tilt at negligible design-effect cost ($\mathrm{DEFF} \le 1.053$).

Finally, Panel D of Figure~\ref{fig:iv_forecasting} and Table~\ref{tab:iv} explain the statistical structure of implied-volatility predictability across the $G = 281$ test sessions. Because NVDA implied volatility spends most calm trading days near its local equilibrium while experiencing sharp upward jumps around earnings and macro tariff announcements followed by rapid multi-day mean reversion, mean-reversion models incur small positive squared-error differentials on a slight majority of calm days in exchange for large squared-error reductions during post-spike volatility reversals. Specifically, at daily 15:00 ET anchors, the sample median loss differential $d_t = e_{\mathrm{model},t}^2 - e_{\mathrm{persist},t}^2$ ($\times 10^4$, or $\text{bps}^2$) is $+2.20$ (5-session moving-block bootstrap \citealp{kunsch1989jackknife} $95\%$ CI $[+0.01, +9.53]$) for HAR-IV and $+0.63$ ($[-0.22, +1.78]$) for Enc-Dec Tier 2 PPS, alongside strongly negative sample means of $-48.8$ and $-44.3$. Because daily loss differentials are heavily left-skewed and serially correlated across sessions, standard \citet{wilcoxon1945individual} signed-rank tests (which test symmetry around zero under independence) should not be over-interpreted as exact tests of the population median; instead, we directly report the sample proportion of sessions favoring persistence ($d_t > 0$) with 5-session moving-block bootstrap $95\%$ CIs: persistence achieves lower squared error on $57.3\%$ ($[50.2\%, 64.8\%]$) of 15:00 ET anchors for HAR-IV and $53.0\%$ ($[47.0\%, 59.4\%]$) for Enc-Dec Tier 2 PPS ($58.7\%$ $[51.2\%, 66.5\%]$ and $60.1\%$ $[54.1\%, 66.5\%]$ on session-mean losses $\bar{d}_g$). Meanwhile, the top 5 post-spike reversal sessions account for $157\%$ (HAR-IV) and $144\%$ (Enc-Dec Tier 2 PPS; $119\%$ census) of net cumulative squared-error gains, while two unpredicted initial jump sessions inflate the parametric batch-means standard error ($t_{\mathrm{sess}} = -0.95, p = 0.34$ for HAR-IV, $t_{\mathrm{sess}} = -0.93, p = 0.35$ for Enc-Dec Tier 2 PPS, and $t_{\mathrm{sess}} = -1.17, p = 0.24$ for Enc-Dec Tier 2 census, tightening to $t_{\mathrm{trim}} = -1.09, p = 0.27$, $t_{\mathrm{trim}} = -1.44, p = 0.15$, and $t_{\mathrm{trim}} = -1.25, p = 0.21$ when those two jump outliers are trimmed). We therefore interpret the $+17.2\%$ to $+21.0\%$ pooled MSE skill as descriptive, episode-concentrated volatility-reversion skill rather than statistically significant uniform daily dominance.

\section{Intraday Activity Clocks Across a Nearly Twentyfold Trough-to-Peak Share Price Expansion}
\label{sec:bars}

To test how high-frequency activity clocks behave across a multi-year valuation expansion \citep{clark1973subordinated, ane2000order, lopez2018advances}, we construct chronological Time Bars ($43\text{ RTH min/bar}$, targeting $9.0\text{ bars/session}$), Volume Bars, and Dollar Bars from DST-corrected 1-minute NVDA bars across $967$ sessions ($2022\text{--}2025$). Across these $967$ sessions, $960$ are full $390$-minute sessions ($09\text{:}30\text{--}15\text{:}59$ ET) and $7$ are NYSE/Nasdaq 13:00 ET early-close holiday sessions ($210$ strict RTH minutes each, for a theoretical strict RTH ceiling of $960 \times 390 + 7 \times 210 = 375{,}870$ minutes). Applying a uniform $09\text{:}30\text{--}15\text{:}59$ ET filter across all $967$ days retains $376{,}838$ 1-minute bars ($8{,}703 = 9 \times 967$ Time Bars), including $972$ post-13:00 ET rows on the $7$ early-close dates; truncating those $7$ early-close sessions strictly at 13:00 ET yields $375{,}866$ strict RTH 1-minute bars ($4$ zero-trade minutes below $375{,}870$) and $8{,}675$ Time Bars ($8.97\text{ bars/session}$). We evaluate all clock diagnostics under both conventions and verify that every kurtosis, autocorrelation, and forecasting metric is numerically similar across the two conventions (Table~\ref{tab:bars} notes).

To eliminate full-sample look-ahead in threshold calibration, the volume threshold $V_d^{\star}$ and dollar threshold $D_d^{\star}$ for trading day $d$ are computed strictly causally from the trailing 60-session rolling mean of daily RTH share volume and dollar turnover up to session $d-1$:
\begin{equation}
    V_d^{\star} = \frac{1}{b^{\star} \times 60} \sum_{j=1}^{60} \mathrm{Vol}_{d-j}^{\mathrm{session}}, \qquad
    D_d^{\star} = \frac{1}{b^{\star} \times 60} \sum_{j=1}^{60} \mathrm{DollarVol}_{d-j}^{\mathrm{session}}, \qquad b^{\star} = \frac{8{,}000}{967} \approx 8.273.
    \label{eq:causal_clock}
\end{equation}
Within each trading session $d$, cumulative share volume or dollar turnover is accumulated minute by minute across $09\text{:}30\text{--}15\text{:}59$ ET, emitting a bar whenever integer multiples of $V_d^{\star}$ or $D_d^{\star}$ are crossed and resetting at the 16:00 ET session close so that overnight gaps never contaminate within-session returns.

To ensure a strictly apples-to-apples comparison across identical intraday time intervals, let $O_{d,1}$ denote the 09:30 ET session open price and $C_{d,1}, \dots, C_{d,K_d}$ denote the bar closing prices on session $d$. Defining the first bar's return as the open-to-close log return $r_{d,1} = \log(C_{d,1}/O_{d,1})$ and subsequent bar returns as $r_{d,j} = \log(C_{d,j}/C_{d,j-1})$ for $j = 2, \dots, K_d$ guarantees that the within-session return sum telescopes identically across all three clocks on every trading day:
\begin{equation}
    \sum_{j=1}^{K_d} r_{d,j} = \log\!\left(\frac{C_{d,1}}{O_{d,1}}\right) + \sum_{j=2}^{K_d} \log\!\left(\frac{C_{d,j}}{C_{d,j-1}}\right) = \log\!\left(\frac{C_{d,K_d}}{O_{d,1}}\right) = \log\!\left(\frac{C_{d,\text{15:59}}}{O_{d,\text{09:30}}}\right).
    \label{eq:telescope}
\end{equation}
Furthermore, to test how much of the excess kurtosis of chronological Time Bars arises from deterministic time-of-day volatility seasonality (the U-shaped intraday variance profile) versus stochastic session-specific activity bursts, we construct a \emph{Diurnally Standardized Time Bars} benchmark: for each intra-session 43-minute slot $b \in \{1, \dots, 9\}$, we estimate the slot return standard deviation $\hat{\sigma}_b$ strictly on the $70\%$ training split ($676$ sessions in $2022\text{--}2024$, where $r_{d,b}/\hat{\sigma}_b$ has unit variance) and evaluate the standardized series across the full $2022\text{--}2025$ sample ($967$ sessions) as a descriptive full-sample standardization benchmark (alongside a strictly out-of-sample check on the last $30\%$ of sessions, $291$ sessions in $2024\text{--}2025$). In Panel A (interior returns), $r_{d,b}/\hat{\sigma}_b$ is renormalized over the full sample to match the interior Time Bar SD ($0.709\%$); in Panel B (full-session returns), multiplying the training-standardized returns $r_{d,b}/\hat{\sigma}_b$ (which have full-sample SD $0.975$ because the $2024\text{--}2025$ test period has slightly lower average intraday volatility than $2022\text{--}2024$) by the unconditional full-sample Time Bar SD ($0.869\%$) yields a realized full-sample SD of $0.847\%$ (note that Fisher excess kurtosis $\kappa = 5.22$ is invariant to global constant rescaling, so rescaling Panel B to $0.869\%$ leaves $\kappa = 5.22$ and $\Delta\kappa = -8.78$ unchanged). In Table~\ref{tab:bars}, we lead in Panel A with Interior-Session Close-to-Close Returns ($j = 2, \dots, K_d$, excluding the opening bar of each session so that opening-bell jumps do not dominate the comparison), and report in Panel B the complete set of Full-Session Intraday Returns ($j = 1, \dots, K_d$, where cumulative log return is identically $0.7059$ and cumulative realized variance $\sum_{d,j} r_{d,j}^2$ is $0.657, 0.624,$ and $0.639$ across Time, Volume, and Dollar clocks). All squared-return autocorrelations $\rho_k(r_t^2)$ are computed strictly over within-session pairs $(r_{d,j}^2, r_{d,j+k}^2)$ so that lag pairs never cross overnight session boundaries, and we report exact sample point differences against both unadjusted Time Bars ($\Delta\kappa_{\mathrm{clock}} = \kappa_{\mathrm{clock}} - \kappa_{\mathrm{Time}}$) and Diurnally Standardized Time Bars ($\Delta\kappa_{\mathrm{diurn}} = \kappa_{\mathrm{clock}} - \kappa_{\mathrm{Diurn}}$) with paired $1{,}000$-replicate moving-block bootstrap \citep{kunsch1989jackknife} $95\%$ confidence intervals resampling contiguous blocks of $b = 10$ trading sessions.

\begin{table}[!htbp]
\centering
\caption{Intraday Microstructure Bar Clocks Constructed from RTH 1-Minute NVDA Bars Across $967$ Sessions ($2022\text{--}2025$).}
\label{tab:bars}
\resizebox{\textwidth}{!}{\IfFileExists{tables/tab_bars.tex}{\input{tables/tab_bars.tex}}{\input{tab_bars.tex}}}
\vspace{2pt}
\begin{minipage}{0.98\textwidth}
\footnotesize \textit{Notes:} Panel A evaluates interior close-to-close returns ($j = 2, \dots, K_d$, $N = \mathrm{Bars} - 967$, excluding the opening bar of each session). Panel B evaluates all within-session bars ($j = 1, \dots, K_d$) including the first bar's open-to-close return $r_{d,1} = \log(C_{d,1}/O_{d,1})$ so that all clocks span the exact same $09\text{:}30\text{--}16\text{:}00$ ET session interval ($\sum r_{d,j} = 0.7059$; integrated variance $\sum r_{d,j}^2 = 0.657, 0.624, 0.639$). ``Time (Diurnally Std.)'' divides each intra-session time-bar slot $b \in \{1,\dots,9\}$ by its $70\%$ training-split ($2022\text{--}2024$) slot standard deviation $\hat{\sigma}_b$ and evaluates across all $967$ sessions ($2022\text{--}2025$; Panel A renormalizes interior returns to $0.709\%$ SD, while Panel B multiplies $r_{d,b}/\hat{\sigma}_b$ by the full-sample Time Bar SD of $0.869\%$, yielding $0.847\%$ full-sample SD; excess kurtosis $\kappa = 5.22$ is invariant to global constant rescaling). ``$\Delta\kappa$ vs.\ Time'' and ``$\Delta\kappa$ vs.\ Diurn.'' report exact sample point differences $\kappa_{\mathrm{clock}} - \kappa_{\mathrm{Time}}$ and $\kappa_{\mathrm{clock}} - \kappa_{\mathrm{Diurn}}$ alongside $95\%$ paired session-block bootstrap CIs ($B=1{,}000$, contiguous block length $b=10$ sessions; \citealp{kunsch1989jackknife}; bootstrap mean differences vs.\ Time are $-3.15$ and $-2.49$ in Panel A and $-11.95$ and $-11.33$ in Panel B, and vs.\ Diurn.\ are $-3.28$ [$p=0.010$] and $-2.62$ [$p<0.001$] in Panel A and $-3.55$ [$p=0.002$] and $-2.93$ [$p<0.001$] in Panel B). Under strict 13:00 ET early-close truncation on the $7$ early-close sessions ($375{,}866$ 1-minute bars, $8{,}675$ Time Bars), results are numerically similar: interior kurtosis is $4.97, 5.07, 1.85,$ and $2.51$ ($\Delta\kappa_{\mathrm{Time}} = -3.12$ and $-2.47$; $\Delta\kappa_{\mathrm{Diurn}} = -3.22$ and $-2.56$) and full-session kurtosis is $13.95, 5.20, 1.69,$ and $2.29$ ($\Delta\kappa_{\mathrm{Time}} = -12.26$ and $-11.65$; $\Delta\kappa_{\mathrm{Diurn}} = -3.51$ and $-2.91$).
\end{minipage}
\end{table}

\begin{figure}[!htbp]
    \centering
    \includegraphics[width=0.92\textwidth]{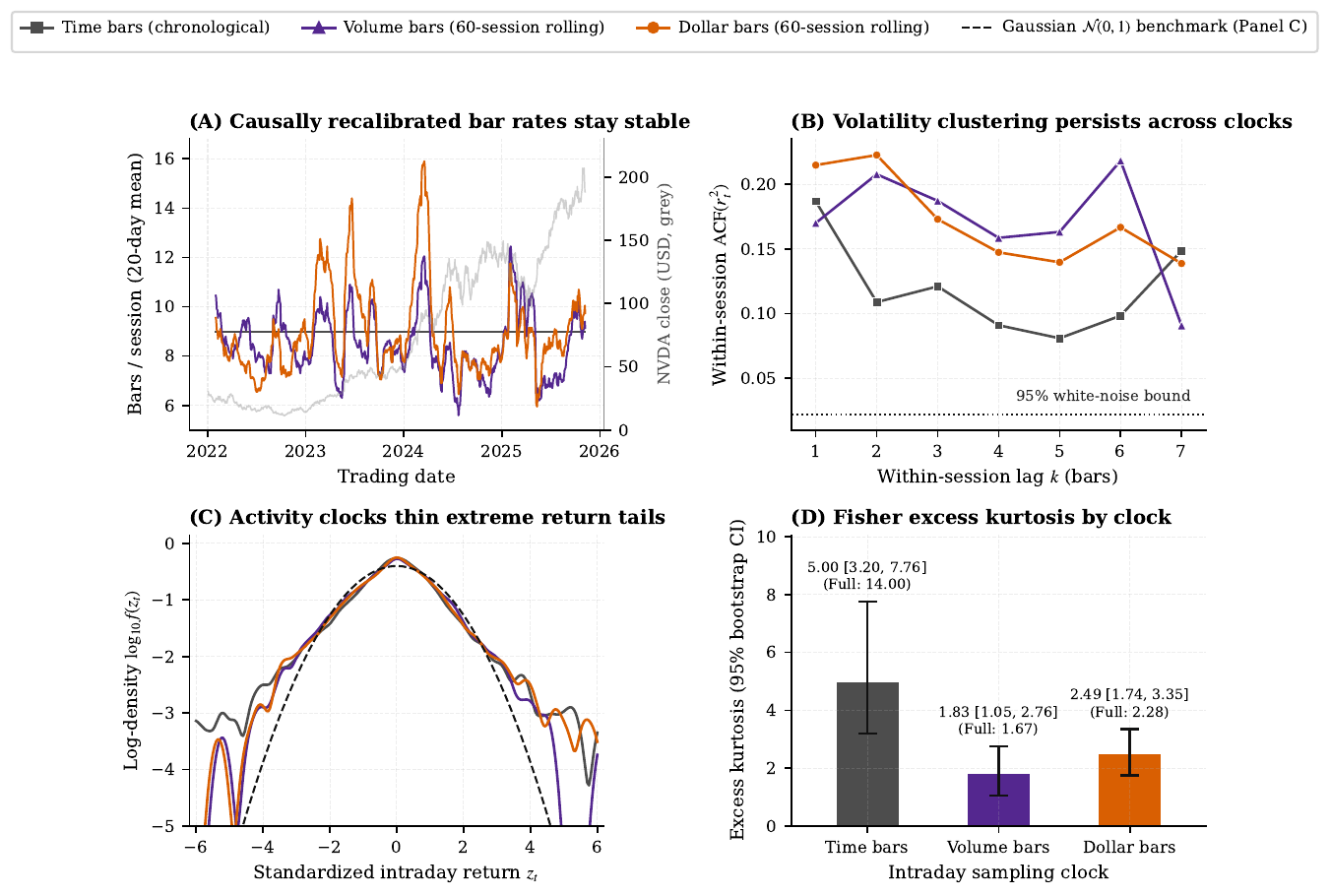}
    \caption{Causally Recalibrated Intraday Activity Clocks on $376{,}838$ 1-Minute NVDA Bars ($2022\text{--}2025$). Panel A: Trailing 20-session mean bar formation rate (bars/session) for Time, Volume, and Dollar clocks overlaid against split-adjusted NVDA share price (grey right axis; $\$29.82$ to $\$188.26$ start-to-end [$6.3\times$], and $\$11.23$ October 2022 trough to $\$212.19$ October 2025 peak [$18.9\times$]). Panel B: Strictly within-session non-crossing squared-return autocorrelation $\mathrm{ACF}(r_t^2)$ across within-session bar lags $k = 1, \dots, 7$ relative to the $95\%$ white-noise bound. Panel C: Log-density $\log_{10} f(z_t)$ of standardized within-session returns versus the $\mathcal{N}(0,1)$ benchmark. Panel D: Fisher excess kurtosis with $95\%$ session-block bootstrap confidence intervals for interior returns and full-session callouts.}
    \label{fig:bar_clocks}
\end{figure}

Table~\ref{tab:bars} and Figure~\ref{fig:bar_clocks} establish three empirical results on intraday activity clocks. First, across the $2022\text{--}2025$ sample, NVDA's split-adjusted share price rose $6.3\times$ start-to-end ($\$29.82$ on January 3, 2022 to $\$188.26$ on November 7, 2025) and $18.9\times$ from its October 2022 trough ($\$11.23$) to its October 2025 peak ($\$212.19$), while the 60-session rolling dollar threshold $D_d^{\star}$ in Equation~\eqref{eq:causal_clock} expanded $3.03\times$ in annual mean from $\$192.7\text{M/bar}$ in 2022 to $\$583.4\text{M/bar}$ in 2025 ($6.41\times$ from the 2022 trough threshold of $\$120.4\text{M/bar}$ to the 2025 peak threshold of $\$771.8\text{M/bar}$). Thus, whereas a static 2022 dollar threshold would fire roughly three to six times more frequently by 2025 (and up to $18.9\times$ more frequently per share traded between the October 2022 price trough and the October 2025 price peak), causal 60-session rolling calibration maintains stable average bar formation rates around the target rate across the four-year span: Time Bars yield $8{,}703$ bars ($9.00\text{ bars/session}$; $8{,}675$ or $8.97\text{ bars/session}$ under strict 13:00 ET early-close truncation), Volume Bars yield $8{,}205$ bars ($8.49\text{ bars/session}$), and Dollar Bars yield $8{,}696$ bars ($8.99\text{ bars/session}$; Panel A of Figure~\ref{fig:bar_clocks}).

Second, comparing chronological Time Bars, Diurnally Standardized Time Bars, and Volume/Dollar activity clocks across Interior-Session Returns (Panel A) and Full-Session Returns (Panel B) separates deterministic time-of-day variance seasonality from dynamic activity subordination. On Interior-Session Close-to-Close Returns excluding the opening bar of each day (Panel A of Table~\ref{tab:bars}), chronological Time Bars exhibit an excess kurtosis of $5.00$ ($[3.20, 7.76]$, $\mathrm{JB} = 8{,}069$), and deterministic diurnal time-of-day standardization leaves interior excess kurtosis virtually unchanged at $5.09$ ($[2.48, 9.44]$, point $\Delta\kappa = +0.09$ $[-1.13, +1.66]$). By contrast, Volume Bars reduce interior excess kurtosis by $63.4\%$ to $1.83$ ($[1.05, 2.76]$, $\mathrm{JB} = 1{,}015$), yielding an exact point difference of $\Delta\kappa = -3.17$ ($95\%$ CI $[-5.99, -1.11]$, $p < 0.001$) relative to Time Bars and $\Delta\kappa_{\mathrm{diurn}} = -3.26$ ($[-6.88, -0.53]$, $p = 0.010$) relative to Diurnally Standardized Time Bars. Similarly, Dollar Bars reduce interior excess kurtosis by $50.2\%$ to $2.49$ ($[1.74, 3.35]$, $\mathrm{JB} = 2{,}011$), yielding $\Delta\kappa = -2.51$ ($[-4.86, -0.75]$, $p = 0.008$) relative to Time Bars and $\Delta\kappa_{\mathrm{diurn}} = -2.60$ ($[-5.66, -0.40]$, $p < 0.001$) relative to Diurnally Standardized Time Bars. On Full-Session Intraday Returns (Panel B of Table~\ref{tab:bars}), where all clocks have similar return standard deviations ($0.847\%\text{--}0.872\%$), chronological Time Bars exhibit an excess kurtosis of $14.00$ ($[4.86, 28.00]$, $\mathrm{JB} = 71{,}434$) because the first 43-minute chronological bar ($09\text{:}30\text{--}10\text{:}12$ ET) absorbs the opening-bell volatility burst into a single observation. Rescaling each time-of-day slot by its training-split standard deviation (Diurnally Standardized Time Bars) lowers full-session excess kurtosis from $14.00$ to $5.22$ ($[2.60, 9.02]$, $\mathrm{JB} = 9{,}932$, point $\Delta\kappa = -8.78$ $[-22.58, +0.73]$), consistent with the interpretation that roughly two-thirds of the full-session Time Bar kurtosis spike reflects the deterministic opening-bell scale jump (though the paired session-block bootstrap interval vs.\ raw Time Bars includes zero); yet dynamic Volume and Dollar clocks go further by adapting to session-specific trading intensity, reducing full-session excess kurtosis to $1.67$ ($[1.04, 2.43]$, $\mathrm{JB} = 952$; $\Delta\kappa = -12.33$ $[-25.97, -3.64]$ vs.\ Time and $\Delta\kappa_{\mathrm{diurn}} = -3.55$ $[-7.40, -0.73], p = 0.002$ vs.\ Diurnally Standardized Time Bars) and $2.28$ ($[1.67, 3.00]$, $\mathrm{JB} = 1{,}884$; $\Delta\kappa = -11.72$ $[-25.30, -3.02]$ vs.\ Time and $\Delta\kappa_{\mathrm{diurn}} = -2.94$ $[-6.28, -0.55], p < 0.001$ vs.\ Diurnally Standardized Time Bars). Evaluating all four clocks on the strictly out-of-sample $30\%$ holdout ($291$ sessions in $2024\text{--}2025$) shows that activity clocks remain lowest out-of-sample: full-session holdout excess kurtosis is $42.27$ for chronological Time Bars ($n = 2{,}619$) and $14.41$ for Diurnally Standardized Time Bars ($\Delta\kappa = -27.86$), compared with $2.34$ for Volume Bars ($n = 2{,}448$, $\Delta\kappa_{\mathrm{diurn}} = -12.07$) and $4.14$ for Dollar Bars ($n = 2{,}538$, $\Delta\kappa_{\mathrm{diurn}} = -10.27$), while interior-session holdout excess kurtosis is $10.81$ for Time Bars ($n = 2{,}328$) and $13.50$ for Diurnally Standardized Time Bars versus $2.68$ for Volume Bars ($n = 2{,}157$) and $4.63$ for Dollar Bars ($n = 2{,}247$).

Third, contrary to folklore claims that activity clocks automatically eliminate conditional heteroskedasticity or create directional return predictability, Panel B of Figure~\ref{fig:bar_clocks} and Table~\ref{tab:bars} show that strictly within-session squared-return autocorrelation remains positive and well above the $95\%$ white-noise bound across within-session lags $k = 1, \dots, 7$ for all clocks: on interior-session returns, $\rho_1(r_t^2) = 0.187$ $[0.128, 0.232]$ for Time Bars ($0.216$ diurnally standardized), $0.170$ $[0.129, 0.216]$ for Volume Bars, and $0.215$ $[0.164, 0.263]$ for Dollar Bars (and $0.157, 0.206, 0.147, 0.199$ on full-session returns). Similarly, training the Encoder-Decoder LSTM (3-seed mean) to forecast next-bar within-session returns yields out-of-sample $R^2_{\mathrm{OOS}}$ near zero ($-0.15\%$ for Time Bars, $-0.14\%$ for Diurnally Standardized Time Bars, $+0.05\%$ for Volume Bars, and $-0.18\%$ for Dollar Bars) and directional accuracy near $50\%$ ($48.6\%, 48.7\%, 50.1\%,$ and $51.0\%$). Thus, the primary econometric benefit of causally recalibrated Volume and Dollar bars is thinning unconditional within-session return tails toward Gaussianity, beyond static diurnal time-of-day scaling, while preserving conditional ARCH volatility clustering, rather than generating directional alpha.

\section{Discussion and Conclusion}
\label{sec:conclusion}

This article has presented an empirical study of probability survey sampling, implied volatility surface compilation, multi-horizon sequence risk forecasting, and activity-driven intraday bar clocks on granular NVDA equity and options data ($2018\text{--}2025$), using prior-session dispersion estimates, 1-session-lagged option features, training-split scaling and latent parameters, and 60-session rolling bar thresholds to prevent look-ahead leakage. For quantitative finance practitioners and financial index providers, our findings establish two practical principles for compiling and modeling derivative risk surfaces. First, the efficiency of an option-chain subsampling scheme depends directly on aligning the sampling design with the target index functional and the stratum weighting of the loss metric: across $432{,}529$ option contracts over $753$ trading days, causal plug-in Equal-Cell allocation ($n_h \propto S_{h,t-1}$) and plug-in Neyman allocation ($n_h \propto N_{h,t} S_{h,t-1}$) with bounded water-filling rounding outperform Simple Random Sampling across all retention budgets $\alpha \in [0.25, 0.80]$ when compiling a neutral equal-weighted implied volatility surface via the Hajek estimator (reducing grid RMSE by $32\%$ and $23\%$ at $\alpha = 0.25$ and by $53\%$ and $31\%$ at $\alpha = 0.80$ while guaranteeing $100\%$ cell coverage), whereas Liquidity PPS reduces RMSE by $86\%$ at $\alpha = 0.60$ when compiling a microstructure liquidity-weighted surface via the Horvitz-Thompson ratio estimator by matching the target weights by construction, but inflates EW surface RMSE by $2.47\text{--}3.90\times$ ($3.91\times$ from rounded table entries) when misapplied across objectives. Second, when splicing multi-year derivative panels across vendor schemas and compressing noisy option aggregates, maintaining full maturity support and pruning noisy activity aggregates are essential: compiling the 2024--2025 option summary series from full-chain OPRA.PILLAR cbbo-1m end-of-day snapshots across all $466$ active trading sessions eliminates maturity-truncation breaks, and removing the 8 noisy auxiliary option aggregates (Tier 2, $D=20$) or replacing them with 3 PCA or OptionVAE latents ($D=23$) restores 5-seed mean IV skill under Liquidity PPS ($\alpha=0.60$) from $+1.8\%$ (Tier 3, $D=28$) to $+11.0\%$ (Tier 2), $+12.3\%$ (PCA replacement), and $+13.6\%$ (OptionVAE replacement, with a descriptive cross-seed SD of $2.8\%$ vs.\ $10.6\%$).

For quantitative risk management and financial machine learning, our longitudinal sequence experiments separate three findings on training-set subsampling: (i) informative Liquidity PPS and Neyman designs at $\alpha = 0.60$ retain $81.2\%$ and $73.7\%$ of extreme realized volatility bursts (versus $59.9\%$ under SRS) at minimal design-weight dispersion ($\mathrm{DEFF} \le 1.053$ under $\gamma_{\mathrm{HT}} = 0.55$); (ii) on the 31-feature Tier 5a benchmark, all three probability designs at $\alpha = 0.60$, including SRS, reduce Encoder-Decoder LSTM training time by $21\%\text{--}22\%$ and satisfy formal Two One-Sided Tests (TOST) equivalence relative to full-census multi-horizon price RMSE ($p_{\mathrm{TOST}} \le 0.002$ within $\pm 0.50\%$) and mean single-model IV RMSE across the 10 seeds ($p_{\mathrm{TOST}} \le 0.020$ within $\pm 2.00\%$); and (iii) direct paired differences between informative longitudinal designs and SRS at $\alpha = 0.60$ are small and statistically insignificant ($p \ge 0.608$), leaving incremental forecasting gains from informative longitudinal selection over SRS unestablished. Out-of-sample evaluation shows that while 1-to-14 hour split-adjusted NVDA prices are not forecastable beyond a random walk with drift among tested models ($|t_{\mathrm{DM}}| \le 0.50$), forward implied volatility exhibits $+21.0\%$ pooled out-of-sample MSE skill under our linear HAR-IV adaptation ($1{,}254\text{ bps}$ vs.\ $1{,}410\text{ bps}$; $+22.0\%$ on strictly future-session targets) and $+17.2\%$ MSE skill under a parsimonious Encoder-Decoder LSTM trained with Liquidity PPS ($1{,}283\text{ bps}$), with the neural sequence model not outperforming the simpler linear HAR-IV benchmark ($t_{\mathrm{sess}} = +0.51, p = 0.61$). Session-clustered Diebold-Mariano inference with \citet{flegal2010batch, flegal2025mcmcse} \texttt{mcmcse} batch-means standard errors against persistence ($p = 0.24\text{--}0.35$) and daily loss-differential diagnostics further show that these descriptive MSE reductions are concentrated in post-shock volatility reversals (top 5 sessions supplying $144\%\text{--}157\%$ of net squared-error gains), whereas persistence achieves lower squared error on slightly more than half of individual sessions ($57.3\%$ $[50.2\%, 64.8\%]$ for HAR-IV and $53.0\%$ $[47.0\%, 59.4\%]$ for Tier 2 PPS at 15:00 ET anchors). Finally, across $376{,}838$ 1-minute bars ($375{,}866$ strict RTH) spanning a $6.3\times$ start-to-end and nearly twentyfold ($18.9\times$) trough-to-peak share price expansion ($3.03\times$ to $6.41\times$ in the rolling dollar threshold $D_d^{\star}$), 60-session rolling Volume and Dollar clocks maintain stable average bar formation rates, significantly reduce full-sample within-session return excess kurtosis relative to both chronological Time Bars and Diurnally Standardized Time Bars ($\Delta\kappa_{\mathrm{diurn}} = -3.26$ $[-6.88, -0.53]$ and $-2.60$ $[-5.66, -0.40]$ on full-sample interior returns; $-3.55$ $[-7.40, -0.73]$ and $-2.94$ $[-6.28, -0.55]$ on full-sample full-session returns), and remain lowest in point excess kurtosis on the 291-session out-of-sample holdout ($\Delta\kappa_{\mathrm{diurn}} = -12.07$ and $-10.27$ on holdout full-session returns), providing a useful pre-processing transformation for intraday tail-risk modeling.

Several limitations of this study point to natural directions for future research. First, our empirical analysis focuses on a single mega-cap underlying asset (NVDA) during an extraordinary valuation and split-adjusted volatility cycle; extending the cross-sectional surface allocation and longitudinal subsampling benchmarks to multi-asset equity panels, index options (SPX/NDX), and rates or commodity derivatives would test how design efficiency scales across broader liquidity and skew regimes. Second, our longitudinal forecasting evaluation is conducted on a single chronological out-of-sample test window ($281$ trading sessions in $2024\text{--}2025$), where implied-volatility forecast gains are concentrated in a small number of post-earnings and macro reversal episodes; longer multi-cycle evaluation windows and regime-switching volatility specifications could sharpen inference on rare jump-reversion dynamics. Third, while our primary subsampling equivalence benchmarks use $S = 10$ paired random seeds, the feature-tier and weight-tempering sensitivity ablations in Table~\ref{tab:ablation} use $S = 5$ seeds due to computational budget constraints, and larger seed ensembles would increase statistical power for detecting small representation and $\gamma_{\mathrm{HT}}$ differences. Finally, while our reproduction scripts and summary statistics are fully documented, raw tick and option-chain redistributions remain subject to third-party licensing terms from commercial market-data vendors (Databento XNAS.ITCH and OPRA.PILLAR).

\section*{Use of AI tools declaration}
The authors declare they have used Artificial Intelligence (AI) tools strictly for language editing, LaTeX typesetting verification, and code debugging assistance during manuscript preparation; all econometric formulations, empirical analyses, and scientific conclusions were designed, executed, and verified by the authors.

\section*{Acknowledgments}
This research was supported by the Undergraduate Research Scholars Program from Graduate Studies \& Research at Salisbury University.

\section*{Conflict of interest}
All authors declare no conflicts of interest in this paper.

\section*{Author contributions}
Deepak Bastola: Conceptualization, Methodology, Software, Formal analysis, Investigation, Writing (original draft, review, and editing), Supervision, Funding acquisition; Eli Prushansky: Data curation, Software, Validation, Formal analysis, Visualization. All authors have read and agreed to the published version of the manuscript.

\section*{Data availability statement}
The longitudinal and cross-sectional datasets and modular reproduction scripts supporting the findings of this study are available from the corresponding author upon reasonable request, subject to third-party data vendor licensing agreements (Databento OPRA.PILLAR and XNAS.ITCH).

\section*{Disclaimer (No Financial Advice)}
This study is conducted strictly for academic and methodological research purposes. Nothing in this manuscript constitutes financial, investment, or options trading advice, nor should any statistical model, simulation, or volatility forecast presented herein be construed as a recommendation or solicitation to buy, sell, or hold any security or derivative instrument.

{\sloppy
\setlength{\bibsep}{3pt plus 0.3ex}
\bibliographystyle{apalike}
\bibliography{references}
}

\end{document}

%% file: tables/tab_sampling.tex
\begin{tabular}{@{}llrrrrrr@{}}
\toprule
 & & \multicolumn{3}{c}{Sample Size \& Weight Dispersion} & \multicolumn{3}{c}{Tail Capture \& Distributional Shift} \\
\cmidrule(lr){3-5}\cmidrule(lr){6-8}
$\alpha$ & Sampling Design & $n$ & $N_{\mathrm{eff}}$ & DEFF & Top-Decile & Max Gap & $W_1(\mathrm{IV})$ \\
 & & & (Kish) & & RV (\%) & (hours) & (bps) \\
\midrule
\multicolumn{8}{@{}l}{\textit{Panel A: High-Compression Budget ($\alpha = 0.40$, $40\%$ of Training Windows)}} \\
$0.40$ & Simple Random (SRS) & 3,656 & 3,656 & 1.000 & 40.7 & 16.7 & 23 \\
 & Liquidity PPS & 3,656 & 3,352 & 1.091 & 64.4 & 30.4 & 571 \\
 & Neyman Stratified & 3,656 & 3,578 & 1.022 & 48.7 & 20.9 & 253 \\
\midrule
\multicolumn{8}{@{}l}{\textit{Panel B: Balanced Pareto Budget ($\alpha = 0.60$, $60\%$ of Training Windows)}} \\
$0.60$ & Simple Random (SRS) & 5,483 & 5,483 & 1.000 & 59.9 & 10.4 & 17 \\
 & Liquidity PPS & 5,483 & 5,205 & 1.053 & 81.2 & 19.0 & 395 \\
 & Neyman Stratified & 5,483 & 5,366 & 1.022 & 73.7 & 13.0 & 250 \\
\midrule
\multicolumn{8}{@{}l}{\textit{Panel C: High-Retention Budget ($\alpha = 0.80$, $80\%$ of Training Windows)}} \\
$0.80$ & Simple Random (SRS) & 7,311 & 7,311 & 1.000 & 80.8 & 6.7 & 10 \\
 & Liquidity PPS & 7,311 & 7,145 & 1.023 & 92.5 & 10.6 & 222 \\
 & Neyman Stratified & 7,311 & 7,199 & 1.016 & 93.0 & 9.2 & 184 \\
\bottomrule
\end{tabular}

%% file: tab_sampling.tex
\begin{tabular}{@{}llrrrrrr@{}}
\toprule
 & & \multicolumn{3}{c}{Sample Size \& Weight Dispersion} & \multicolumn{3}{c}{Tail Capture \& Distributional Shift} \\
\cmidrule(lr){3-5}\cmidrule(lr){6-8}
$\alpha$ & Sampling Design & $n$ & $N_{\mathrm{eff}}$ & DEFF & Top-Decile & Max Gap & $W_1(\mathrm{IV})$ \\
 & & & (Kish) & & RV (\%) & (hours) & (bps) \\
\midrule
\multicolumn{8}{@{}l}{\textit{Panel A: High-Compression Budget ($\alpha = 0.40$, $40\%$ of Training Windows)}} \\
$0.40$ & Simple Random (SRS) & 3,656 & 3,656 & 1.000 & 40.7 & 16.7 & 23 \\
 & Liquidity PPS & 3,656 & 3,352 & 1.091 & 64.4 & 30.4 & 571 \\
 & Neyman Stratified & 3,656 & 3,578 & 1.022 & 48.7 & 20.9 & 253 \\
\midrule
\multicolumn{8}{@{}l}{\textit{Panel B: Balanced Pareto Budget ($\alpha = 0.60$, $60\%$ of Training Windows)}} \\
$0.60$ & Simple Random (SRS) & 5,483 & 5,483 & 1.000 & 59.9 & 10.4 & 17 \\
 & Liquidity PPS & 5,483 & 5,205 & 1.053 & 81.2 & 19.0 & 395 \\
 & Neyman Stratified & 5,483 & 5,366 & 1.022 & 73.7 & 13.0 & 250 \\
\midrule
\multicolumn{8}{@{}l}{\textit{Panel C: High-Retention Budget ($\alpha = 0.80$, $80\%$ of Training Windows)}} \\
$0.80$ & Simple Random (SRS) & 7,311 & 7,311 & 1.000 & 80.8 & 6.7 & 10 \\
 & Liquidity PPS & 7,311 & 7,145 & 1.023 & 92.5 & 10.6 & 222 \\
 & Neyman Stratified & 7,311 & 7,199 & 1.016 & 93.0 & 9.2 & 184 \\
\bottomrule
\end{tabular}

%% file: tables/tab_cross.tex
\begin{tabular}{@{}llrrrrccc@{}}
\toprule
 & & \multicolumn{4}{c}{Surface Grid RMSE (bps of IV)} & \multicolumn{3}{c}{RMSE Ratio vs.\ SRS [95\% CI]} \\
\cmidrule(lr){3-6}\cmidrule(lr){7-9}
Target Surface & $\alpha$ & SRS & Neyman & Eq.-Cell & PPS & Neyman / SRS & Eq.-Cell / SRS & PPS / SRS \\
 & & & ($N_h S_{h,t-1}$) & ($S_{h,t-1}$) & ($w_{i,t}$) & & & \\
\midrule
\multicolumn{9}{@{}l}{\textit{Panel A: Neutral Equal-Weighted (EW) Target Surface (Hajek Estimator)}} \\
Equal-Weighted & $0.25$ & 124.0 & 94.9 & 84.8 & 306.1 & 0.77 [0.74, 0.79] & 0.68 [0.66, 0.70] & 2.47 [2.37, 2.59] \\
 & $0.40$ & 81.3 & 66.0 & 55.3 & 249.4 & 0.81 [0.79, 0.83] & 0.68 [0.66, 0.70] & 3.07 [2.94, 3.21] \\
 & $0.60$ & 53.8 & 41.0 & 31.3 & 186.3 & 0.76 [0.73, 0.80] & 0.58 [0.56, 0.60] & 3.46 [3.28, 3.66] \\
 & $0.80$ & 32.0 & 22.2 & 15.1 & 125.1 & 0.69 [0.66, 0.73] & 0.47 [0.45, 0.49] & 3.90 [3.69, 4.15] \\
\midrule
\multicolumn{9}{@{}l}{\textit{Panel B: Liquidity-Weighted (LW) Target Surface (Horvitz-Thompson Ratio Estimator)}} \\
Liquidity-Weighted & $0.25$ & 179.9 & 151.4 & 140.8 & 82.7 & 0.84 [0.82, 0.87] & 0.78 [0.76, 0.81] & 0.46 [0.42, 0.50] \\
 & $0.40$ & 129.3 & 111.5 & 98.3 & 36.6 & 0.86 [0.82, 0.90] & 0.76 [0.73, 0.79] & 0.28 [0.26, 0.31] \\
 & $0.60$ & 84.4 & 71.6 & 58.5 & 11.7 & 0.85 [0.82, 0.88] & 0.69 [0.66, 0.72] & 0.14 [0.13, 0.15] \\
 & $0.80$ & 53.2 & 39.2 & 31.9 & 2.7 & 0.74 [0.70, 0.78] & 0.60 [0.56, 0.64] & 0.05 [0.05, 0.06] \\
\bottomrule
\end{tabular}

%% file: tab_cross.tex
\begin{tabular}{@{}llrrrrccc@{}}
\toprule
 & & \multicolumn{4}{c}{Surface Grid RMSE (bps of IV)} & \multicolumn{3}{c}{RMSE Ratio vs.\ SRS [95\% CI]} \\
\cmidrule(lr){3-6}\cmidrule(lr){7-9}
Target Surface & $\alpha$ & SRS & Neyman & Eq.-Cell & PPS & Neyman / SRS & Eq.-Cell / SRS & PPS / SRS \\
 & & & ($N_h S_{h,t-1}$) & ($S_{h,t-1}$) & ($w_{i,t}$) & & & \\
\midrule
\multicolumn{9}{@{}l}{\textit{Panel A: Neutral Equal-Weighted (EW) Target Surface (Hajek Estimator)}} \\
Equal-Weighted & $0.25$ & 124.0 & 94.9 & 84.8 & 306.1 & 0.77 [0.74, 0.79] & 0.68 [0.66, 0.70] & 2.47 [2.37, 2.59] \\
 & $0.40$ & 81.3 & 66.0 & 55.3 & 249.4 & 0.81 [0.79, 0.83] & 0.68 [0.66, 0.70] & 3.07 [2.94, 3.21] \\
 & $0.60$ & 53.8 & 41.0 & 31.3 & 186.3 & 0.76 [0.73, 0.80] & 0.58 [0.56, 0.60] & 3.46 [3.28, 3.66] \\
 & $0.80$ & 32.0 & 22.2 & 15.1 & 125.1 & 0.69 [0.66, 0.73] & 0.47 [0.45, 0.49] & 3.90 [3.69, 4.15] \\
\midrule
\multicolumn{9}{@{}l}{\textit{Panel B: Liquidity-Weighted (LW) Target Surface (Horvitz-Thompson Ratio Estimator)}} \\
Liquidity-Weighted & $0.25$ & 179.9 & 151.4 & 140.8 & 82.7 & 0.84 [0.82, 0.87] & 0.78 [0.76, 0.81] & 0.46 [0.42, 0.50] \\
 & $0.40$ & 129.3 & 111.5 & 98.3 & 36.6 & 0.86 [0.82, 0.90] & 0.76 [0.73, 0.79] & 0.28 [0.26, 0.31] \\
 & $0.60$ & 84.4 & 71.6 & 58.5 & 11.7 & 0.85 [0.82, 0.88] & 0.69 [0.66, 0.72] & 0.14 [0.13, 0.15] \\
 & $0.80$ & 53.2 & 39.2 & 31.9 & 2.7 & 0.74 [0.70, 0.78] & 0.60 [0.56, 0.64] & 0.05 [0.05, 0.06] \\
\bottomrule
\end{tabular}

%% file: tables/tab_subsampling.tex
\begin{tabular}{@{}llrcrrrcrrr@{}}
\toprule
 & & \multicolumn{4}{c}{Price RMSE (Census \$4.156, $\Delta_{\mathrm{eq}}{=}\pm 0.50\%$)} & \multicolumn{4}{c}{IV RMSE (Census 1377 bps, $\Delta_{\mathrm{eq}}{=}\pm 2.00\%$)} & Train \\
\cmidrule(lr){3-6}\cmidrule(lr){7-10}
$\alpha$ & Sampling Design & $\Delta$\% & 90\% TOST CI & $p_{\mathrm{diff}}$ & $p_{\mathrm{TOST}}$ & $\Delta$\% & 90\% TOST CI & $p_{\mathrm{diff}}$ & $p_{\mathrm{TOST}}$ & time $\Delta$\% \\
\midrule
\multicolumn{11}{@{}l}{\textit{Panel A: High-Compression Subsampling ($\alpha = 0.40$, $n = 3{,}656$)}} \\
$0.40$ & Simple Random (SRS) & $+0.50$ & $[-0.04, +1.03]$ & 0.123 & 0.495 & $+0.82$ & $[-0.43, +2.07]$ & 0.258 & 0.059 & $-36$ \\
 & Liquidity PPS & $+0.34$ & $[+0.01, +0.68]$ & 0.093 & 0.206 & $+2.53$ & $[+1.05, +4.01]$ & 0.012 & 0.736 & $-37$ \\
 & Neyman Stratified & $+0.29$ & $[-0.03, +0.61]$ & 0.130 & 0.130 & $+1.14$ & $[+0.26, +2.02]$ & 0.042 & 0.054 & $-29$ \\
\midrule
\multicolumn{11}{@{}l}{\textit{Panel B: Pareto-Optimal Subsampling ($\alpha = 0.60$, $n = 5{,}483$)}} \\
$0.60$ & Simple Random (SRS) & $+0.11$ & $[-0.07, +0.30]$ & 0.277 & 0.002 & $-0.22$ & $[-1.58, +1.14]$ & 0.775 & 0.020 & $-22$ \\
 & Liquidity PPS & $+0.14$ & $[-0.04, +0.31]$ & 0.192 & 0.002 & $+0.29$ & $[-0.82, +1.40]$ & 0.642 & 0.010 & $-21$ \\
 & Neyman Stratified & $+0.12$ & $[+0.02, +0.22]$ & 0.048 & $<0.001$ & $-0.13$ & $[-1.48, +1.21]$ & 0.860 & 0.016 & $-22$ \\
\midrule
\multicolumn{11}{@{}l}{\textit{Panel C: High-Retention Subsampling ($\alpha = 0.80$, $n = 7{,}311$)}} \\
$0.80$ & Simple Random (SRS) & $+0.08$ & $[-0.06, +0.21]$ & 0.314 & $<0.001$ & $-0.22$ & $[-1.54, +1.09]$ & 0.761 & 0.018 & $-13$ \\
 & Liquidity PPS & $+0.15$ & $[-0.01, +0.30]$ & 0.114 & 0.001 & $+0.42$ & $[-0.92, +1.76]$ & 0.579 & 0.030 & $-4$ \\
 & Neyman Stratified & $+0.15$ & $[-0.03, +0.34]$ & 0.165 & 0.004 & $+0.85$ & $[-0.75, +2.45]$ & 0.355 & 0.110 & $-20$ \\
\bottomrule
\end{tabular}

%% file: tab_subsampling.tex
\begin{tabular}{@{}llrcrrrcrrr@{}}
\toprule
 & & \multicolumn{4}{c}{Price RMSE (Census \$4.156, $\Delta_{\mathrm{eq}}{=}\pm 0.50\%$)} & \multicolumn{4}{c}{IV RMSE (Census 1377 bps, $\Delta_{\mathrm{eq}}{=}\pm 2.00\%$)} & Train \\
\cmidrule(lr){3-6}\cmidrule(lr){7-10}
$\alpha$ & Sampling Design & $\Delta$\% & 90\% TOST CI & $p_{\mathrm{diff}}$ & $p_{\mathrm{TOST}}$ & $\Delta$\% & 90\% TOST CI & $p_{\mathrm{diff}}$ & $p_{\mathrm{TOST}}$ & time $\Delta$\% \\
\midrule
\multicolumn{11}{@{}l}{\textit{Panel A: High-Compression Subsampling ($\alpha = 0.40$, $n = 3{,}656$)}} \\
$0.40$ & Simple Random (SRS) & $+0.50$ & $[-0.04, +1.03]$ & 0.123 & 0.495 & $+0.82$ & $[-0.43, +2.07]$ & 0.258 & 0.059 & $-36$ \\
 & Liquidity PPS & $+0.34$ & $[+0.01, +0.68]$ & 0.093 & 0.206 & $+2.53$ & $[+1.05, +4.01]$ & 0.012 & 0.736 & $-37$ \\
 & Neyman Stratified & $+0.29$ & $[-0.03, +0.61]$ & 0.130 & 0.130 & $+1.14$ & $[+0.26, +2.02]$ & 0.042 & 0.054 & $-29$ \\
\midrule
\multicolumn{11}{@{}l}{\textit{Panel B: Pareto-Optimal Subsampling ($\alpha = 0.60$, $n = 5{,}483$)}} \\
$0.60$ & Simple Random (SRS) & $+0.11$ & $[-0.07, +0.30]$ & 0.277 & 0.002 & $-0.22$ & $[-1.58, +1.14]$ & 0.775 & 0.020 & $-22$ \\
 & Liquidity PPS & $+0.14$ & $[-0.04, +0.31]$ & 0.192 & 0.002 & $+0.29$ & $[-0.82, +1.40]$ & 0.642 & 0.010 & $-21$ \\
 & Neyman Stratified & $+0.12$ & $[+0.02, +0.22]$ & 0.048 & $<0.001$ & $-0.13$ & $[-1.48, +1.21]$ & 0.860 & 0.016 & $-22$ \\
\midrule
\multicolumn{11}{@{}l}{\textit{Panel C: High-Retention Subsampling ($\alpha = 0.80$, $n = 7{,}311$)}} \\
$0.80$ & Simple Random (SRS) & $+0.08$ & $[-0.06, +0.21]$ & 0.314 & $<0.001$ & $-0.22$ & $[-1.54, +1.09]$ & 0.761 & 0.018 & $-13$ \\
 & Liquidity PPS & $+0.15$ & $[-0.01, +0.30]$ & 0.114 & 0.001 & $+0.42$ & $[-0.92, +1.76]$ & 0.579 & 0.030 & $-4$ \\
 & Neyman Stratified & $+0.15$ & $[-0.03, +0.34]$ & 0.165 & 0.004 & $+0.85$ & $[-0.75, +2.45]$ & 0.355 & 0.110 & $-20$ \\
\bottomrule
\end{tabular}

%% file: tables/tab_price.tex
\begin{tabular}{@{}lrrrrrrr@{}}
\toprule
 & \multicolumn{4}{c}{Out-of-Sample Price RMSE (\$)} & \multicolumn{3}{c}{Skill \& Direction} \\
\cmidrule(lr){2-5}\cmidrule(lr){6-8}
Forecasting Model & Pooled & $h{=}1$ & $h{=}7$ & $h{=}14$ & FSS & DM & $\mathrm{DA}_{7\mathrm{h}}$ \\
 & & & & & (\%) & $t_{h=7}$ & (\%) \\
\midrule
\multicolumn{8}{@{}l}{\textit{Panel A: Martingale, Drift, and Linear Regularized Benchmarks}} \\
Random walk (no-change) & 4.148 & 1.516 & 4.013 & 5.636 & $+0.00$ & -- & -- \\
Historical drift & 4.141 & 1.515 & 4.008 & 5.624 & $+0.30$ & $-0.47$ & 55.17 \\
Linear Ridge ($D=31$, all features) & 4.207 & 1.517 & 4.060 & 5.751 & $-2.90$ & $+1.27$ & 50.64 \\
Linear Ridge ($D=16$, price features) & 4.145 & 1.515 & 4.010 & 5.631 & $+0.13$ & $-0.32$ & 55.17 \\
\midrule
\multicolumn{8}{@{}l}{\textit{Panel B: Encoder-Decoder LSTM Across Survey Designs (Tier 5a, $D=31$, 10-Seed Mean $\pm$ SD)}} \\
Enc-Dec, Full census ($\alpha=1.0$) & $4.156 \pm 0.008$ & 1.522 & 4.019 & 5.648 & $-0.39 \pm 0.41$ & $+0.24$ & $55.0 \pm 0.7$ \\
Enc-Dec, SRS ($\alpha=0.6$) & $4.160 \pm 0.014$ & 1.524 & 4.022 & 5.657 & $-0.62 \pm 0.68$ & $+0.43$ & $54.7 \pm 0.8$ \\
Enc-Dec, Liquidity PPS ($\alpha=0.6$) & $4.161 \pm 0.009$ & 1.523 & 4.024 & 5.657 & $-0.66 \pm 0.45$ & $+0.50$ & $54.8 \pm 0.7$ \\
Enc-Dec, Neyman ($\alpha=0.6$) & $4.161 \pm 0.010$ & 1.523 & 4.022 & 5.659 & $-0.63 \pm 0.50$ & $+0.47$ & $55.0 \pm 0.5$ \\
\bottomrule
\end{tabular}

%% file: tab_price.tex
\begin{tabular}{@{}lrrrrrrr@{}}
\toprule
 & \multicolumn{4}{c}{Out-of-Sample Price RMSE (\$)} & \multicolumn{3}{c}{Skill \& Direction} \\
\cmidrule(lr){2-5}\cmidrule(lr){6-8}
Forecasting Model & Pooled & $h{=}1$ & $h{=}7$ & $h{=}14$ & FSS & DM & $\mathrm{DA}_{7\mathrm{h}}$ \\
 & & & & & (\%) & $t_{h=7}$ & (\%) \\
\midrule
\multicolumn{8}{@{}l}{\textit{Panel A: Martingale, Drift, and Linear Regularized Benchmarks}} \\
Random walk (no-change) & 4.148 & 1.516 & 4.013 & 5.636 & $+0.00$ & -- & -- \\
Historical drift & 4.141 & 1.515 & 4.008 & 5.624 & $+0.30$ & $-0.47$ & 55.17 \\
Linear Ridge ($D=31$, all features) & 4.207 & 1.517 & 4.060 & 5.751 & $-2.90$ & $+1.27$ & 50.64 \\
Linear Ridge ($D=16$, price features) & 4.145 & 1.515 & 4.010 & 5.631 & $+0.13$ & $-0.32$ & 55.17 \\
\midrule
\multicolumn{8}{@{}l}{\textit{Panel B: Encoder-Decoder LSTM Across Survey Designs (Tier 5a, $D=31$, 10-Seed Mean $\pm$ SD)}} \\
Enc-Dec, Full census ($\alpha=1.0$) & $4.156 \pm 0.008$ & 1.522 & 4.019 & 5.648 & $-0.39 \pm 0.41$ & $+0.24$ & $55.0 \pm 0.7$ \\
Enc-Dec, SRS ($\alpha=0.6$) & $4.160 \pm 0.014$ & 1.524 & 4.022 & 5.657 & $-0.62 \pm 0.68$ & $+0.43$ & $54.7 \pm 0.8$ \\
Enc-Dec, Liquidity PPS ($\alpha=0.6$) & $4.161 \pm 0.009$ & 1.523 & 4.024 & 5.657 & $-0.66 \pm 0.45$ & $+0.50$ & $54.8 \pm 0.7$ \\
Enc-Dec, Neyman ($\alpha=0.6$) & $4.161 \pm 0.010$ & 1.523 & 4.022 & 5.659 & $-0.63 \pm 0.50$ & $+0.47$ & $55.0 \pm 0.5$ \\
\bottomrule
\end{tabular}

%% file: tables/tab_iv.tex
\begin{tabular}{@{}lrrrrrrcr@{}}
\toprule
 & \multicolumn{3}{c}{Out-of-Sample Accuracy} & \multicolumn{3}{c}{Session-Clustered DM ($G{=}281$)} & \multicolumn{2}{c}{Daily 15:00 ET Loss Diff.\ $d_t$} \\
\cmidrule(lr){2-4}\cmidrule(lr){5-7}\cmidrule(lr){8-9}
Volatility Forecaster & RMSE & Pooled & Fut.-Sess. & vs.\ Persist. & Trim-2 & vs.\ HAR-IV & Median $d_t$ & Persist.\ Win \% \\
 & (bps) & Skill (\%) & Skill (\%) & $t_{\mathrm{sess}}$ ($p$) & $t_{\mathrm{trim}}$ ($p$) & $t_{\mathrm{sess}}$ ($p$) & [95\% Boot.\ CI] & [95\% Boot.\ CI] \\
\midrule
\multicolumn{9}{@{}l}{\textit{Panel A: Persistence \& Multi-Horizon Linear Autoregression ($N_{\mathrm{test}} = 1{,}963 \times 14 = 27{,}482$ Pairs)}} \\
Persistence (last EOD IV) & 1,410 & $0.0$ & $0.0$ & -- & -- & $+0.95$ (0.34) & -- & -- \\
HAR-IV (multi-horizon linear) & 1,254 & $+21.0$ & $+22.0$ & $-0.95$ (0.34) & $-1.09$ (0.27) & -- & $+2.20$ $[+0.01, +9.53]$ & 57.3 [50.2, 64.8] \\
\midrule
\multicolumn{9}{@{}l}{\textit{Panel B: Encoder-Decoder LSTM (Seed-Ensemble Forecasts Across Tiers and Survey Designs)}} \\
Enc-Dec, Tier 2 (IV/skew/VRP), PPS $\alpha{=}0.6$ & 1,283 & $+17.2$ & $+18.2$ & $-0.93$ (0.35) & $-1.44$ (0.15) & $+0.51$ (0.61) & $+0.63$ $[-0.22, +1.78]$ & 53.0 [47.0, 59.4] \\
Enc-Dec, Tier 2 (IV/skew/VRP), Full census & 1,286 & $+16.9$ & $+17.9$ & $-1.17$ (0.24) & $-1.25$ (0.21) & $+0.46$ (0.64) & $+1.85$ $[+0.72, +3.91]$ & 60.5 [54.4, 67.6] \\
Enc-Dec, Tier 5a (All + VAE), Neyman $\alpha{=}0.6$ & 1,339 & $+9.9$ & $+10.8$ & $-0.62$ (0.54) & $-0.85$ (0.40) & $+1.42$ (0.16) & $+5.43$ $[+2.84, +12.08]$ & 66.9 [60.9, 73.3] \\
Enc-Dec, Tier 3 (All options), Full census & 1,344 & $+9.2$ & $+10.1$ & $-0.63$ (0.53) & $-0.74$ (0.46) & $+1.35$ (0.18) & $+5.75$ $[+2.02, +12.89]$ & 63.0 [56.9, 69.0] \\
Enc-Dec, Tier 5a (All + VAE), Full census & 1,346 & $+8.9$ & $+9.7$ & $-0.58$ (0.56) & $-0.82$ (0.41) & $+1.48$ (0.14) & $+6.23$ $[+2.40, +13.06]$ & 65.1 [58.7, 71.2] \\
Enc-Dec, Tier 1 (Price/technical only) & 1,411 & $-0.1$ & $-0.1$ & $+0.44$ (0.66) & $-0.27$ (0.79) & $+0.95$ (0.34) & $+0.05$ $[-0.14, +0.14]$ & 52.0 [45.2, 58.0] \\
\midrule
\multicolumn{9}{@{}l}{\textit{Panel C: Pure Daily Session-Ahead HAR-IV Benchmark (1- and 2-Session-Ahead EOD IV on Unique Sessions)}} \\
Daily HAR-IV, 1-Session Ahead ($\mathrm{IV}_{d+1}$, $G{=}281$) & 1,085 & $+17.4$ & $+17.4$ & $-0.91$ (0.36) & $-1.23$ (0.22) & -- & $+1.65$ $[+0.16, +4.14]$ & 58.7 [52.0, 65.5] \\
Daily HAR-IV, 2-Session Ahead ($\mathrm{IV}_{d+2}$, $G{=}280$) & 1,258 & $+22.9$ & $+22.9$ & $-1.44$ (0.15) & $-1.25$ (0.21) & -- & $+1.90$ $[+0.09, +7.14]$ & 57.3 [50.9, 64.4] \\
\bottomrule
\end{tabular}

%% file: tab_iv.tex
\begin{tabular}{@{}lrrrrrrcr@{}}
\toprule
 & \multicolumn{3}{c}{Out-of-Sample Accuracy} & \multicolumn{3}{c}{Session-Clustered DM ($G{=}281$)} & \multicolumn{2}{c}{Daily 15:00 ET Loss Diff.\ $d_t$} \\
\cmidrule(lr){2-4}\cmidrule(lr){5-7}\cmidrule(lr){8-9}
Volatility Forecaster & RMSE & Pooled & Fut.-Sess. & vs.\ Persist. & Trim-2 & vs.\ HAR-IV & Median $d_t$ & Persist.\ Win \% \\
 & (bps) & Skill (\%) & Skill (\%) & $t_{\mathrm{sess}}$ ($p$) & $t_{\mathrm{trim}}$ ($p$) & $t_{\mathrm{sess}}$ ($p$) & [95\% Boot.\ CI] & [95\% Boot.\ CI] \\
\midrule
\multicolumn{9}{@{}l}{\textit{Panel A: Persistence \& Multi-Horizon Linear Autoregression ($N_{\mathrm{test}} = 1{,}963 \times 14 = 27{,}482$ Pairs)}} \\
Persistence (last EOD IV) & 1,410 & $0.0$ & $0.0$ & -- & -- & $+0.95$ (0.34) & -- & -- \\
HAR-IV (multi-horizon linear) & 1,254 & $+21.0$ & $+22.0$ & $-0.95$ (0.34) & $-1.09$ (0.27) & -- & $+2.20$ $[+0.01, +9.53]$ & 57.3 [50.2, 64.8] \\
\midrule
\multicolumn{9}{@{}l}{\textit{Panel B: Encoder-Decoder LSTM (Seed-Ensemble Forecasts Across Tiers and Survey Designs)}} \\
Enc-Dec, Tier 2 (IV/skew/VRP), PPS $\alpha{=}0.6$ & 1,283 & $+17.2$ & $+18.2$ & $-0.93$ (0.35) & $-1.44$ (0.15) & $+0.51$ (0.61) & $+0.63$ $[-0.22, +1.78]$ & 53.0 [47.0, 59.4] \\
Enc-Dec, Tier 2 (IV/skew/VRP), Full census & 1,286 & $+16.9$ & $+17.9$ & $-1.17$ (0.24) & $-1.25$ (0.21) & $+0.46$ (0.64) & $+1.85$ $[+0.72, +3.91]$ & 60.5 [54.4, 67.6] \\
Enc-Dec, Tier 5a (All + VAE), Neyman $\alpha{=}0.6$ & 1,339 & $+9.9$ & $+10.8$ & $-0.62$ (0.54) & $-0.85$ (0.40) & $+1.42$ (0.16) & $+5.43$ $[+2.84, +12.08]$ & 66.9 [60.9, 73.3] \\
Enc-Dec, Tier 3 (All options), Full census & 1,344 & $+9.2$ & $+10.1$ & $-0.63$ (0.53) & $-0.74$ (0.46) & $+1.35$ (0.18) & $+5.75$ $[+2.02, +12.89]$ & 63.0 [56.9, 69.0] \\
Enc-Dec, Tier 5a (All + VAE), Full census & 1,346 & $+8.9$ & $+9.7$ & $-0.58$ (0.56) & $-0.82$ (0.41) & $+1.48$ (0.14) & $+6.23$ $[+2.40, +13.06]$ & 65.1 [58.7, 71.2] \\
Enc-Dec, Tier 1 (Price/technical only) & 1,411 & $-0.1$ & $-0.1$ & $+0.44$ (0.66) & $-0.27$ (0.79) & $+0.95$ (0.34) & $+0.05$ $[-0.14, +0.14]$ & 52.0 [45.2, 58.0] \\
\midrule
\multicolumn{9}{@{}l}{\textit{Panel C: Pure Daily Session-Ahead HAR-IV Benchmark (1- and 2-Session-Ahead EOD IV on Unique Sessions)}} \\
Daily HAR-IV, 1-Session Ahead ($\mathrm{IV}_{d+1}$, $G{=}281$) & 1,085 & $+17.4$ & $+17.4$ & $-0.91$ (0.36) & $-1.23$ (0.22) & -- & $+1.65$ $[+0.16, +4.14]$ & 58.7 [52.0, 65.5] \\
Daily HAR-IV, 2-Session Ahead ($\mathrm{IV}_{d+2}$, $G{=}280$) & 1,258 & $+22.9$ & $+22.9$ & $-1.44$ (0.15) & $-1.25$ (0.21) & -- & $+1.90$ $[+0.09, +7.14]$ & 57.3 [50.9, 64.4] \\
\bottomrule
\end{tabular}

%% file: tables/tab_ablation.tex
\begin{tabular}{@{}lcrrrr@{}}
\toprule
 & & \multicolumn{2}{c}{Full Census ($\alpha=1.00$) / Neyman ($\alpha=0.60$)} & \multicolumn{2}{c}{Liquidity PPS ($\alpha=0.60$)} \\
\cmidrule(lr){3-4}\cmidrule(lr){5-6}
Feature Representation / Weighting Arm & $D$ & Price FSS (\%) & IV Skill (\%) & Price FSS (\%) & IV Skill (\%) \\
\midrule
\multicolumn{6}{@{}l}{\textit{Panel A: Nested Feature \& Latent Representation Tiers ($\gamma_{\mathrm{HT}} = 0.55$; Left: Full Census, Right: PPS $\alpha=0.60$)}} \\
Tier 1: Price \& technicals only & 16 & $+0.17 \pm 0.10$ & $-0.1 \pm 0.2$ & $+0.06 \pm 0.13$ & $-0.3 \pm 0.3$ \\
Tier 2: + IV level, skew, $\Delta\mathrm{IV}_{1\mathrm{d}}$, VRP proxy & 20 & $-0.04 \pm 0.50$ & $+13.4 \pm 3.7$ & $-0.42 \pm 0.41$ & $+11.0 \pm 10.6$ \\
Tier 3: + 8 option chain activity aggregates & 28 & $-0.44 \pm 1.31$ & $+3.9 \pm 5.1$ & $-0.58 \pm 0.85$ & $+1.8 \pm 3.5$ \\
Tier 4a: + 3 PCA option-state latents (Augmentation) & 31 & $-0.26 \pm 0.58$ & $+7.7 \pm 3.1$ & $-1.08 \pm 0.38$ & $+7.7 \pm 1.1$ \\
Tier 4b: + 3 PCA option-state latents (Replacement) & 23 & $-0.07 \pm 0.21$ & $+13.0 \pm 5.0$ & $-0.40 \pm 0.69$ & $+12.3 \pm 5.3$ \\
Tier 5a: + 3 OptionVAE latents (Augmentation) & 31 & $-0.29 \pm 0.30$ & $+4.4 \pm 5.4$ & $-0.92 \pm 0.48$ & $+3.0 \pm 5.1$ \\
Tier 5b: + 3 OptionVAE latents (Replacement) & 23 & $+0.13 \pm 0.30$ & $+11.8 \pm 6.3$ & $-0.32 \pm 0.45$ & $+13.6 \pm 2.8$ \\
\midrule
\multicolumn{6}{@{}l}{\textit{Panel B: Weight-Tempering Sensitivity at $\alpha = 0.60$ (Left: Tier 5a Neyman / PPS; Right: Tier 5b Replacement PPS)}} \\
Unweighted subsampling ($\gamma_{\mathrm{HT}} = 0.00$, Tier 5a / 5b) & 31 / 23 & $-1.44 \pm 1.32$ & $+1.8 \pm 5.7$ & $-0.97 \pm 0.57$ & $+0.1 \pm 6.0$ \;($+12.5 \pm 4.1$) \\
Tempered Inverse-Probability Weighting ($\gamma_{\mathrm{HT}} = 0.55$, Tier 5a / 5b) & 31 / 23 & $-0.54 \pm 0.48$ & $+3.2 \pm 5.1$ & $-0.92 \pm 0.48$ & $+3.0 \pm 5.1$ \;($+13.6 \pm 2.8$) \\
Self-Normalized Inverse-Probability Weighting ($\gamma_{\mathrm{HT}} = 1.00$, H\'{a}jek Ratio IPW, Tier 5a / 5b) & 31 / 23 & $-0.35 \pm 0.44$ & $+3.2 \pm 4.0$ & $-0.60 \pm 0.32$ & $+5.5 \pm 5.2$ \;($+14.0 \pm 2.6$) \\
\bottomrule
\end{tabular}

%% file: tab_ablation.tex
\begin{tabular}{@{}lcrrrr@{}}
\toprule
 & & \multicolumn{2}{c}{Full Census ($\alpha=1.00$) / Neyman ($\alpha=0.60$)} & \multicolumn{2}{c}{Liquidity PPS ($\alpha=0.60$)} \\
\cmidrule(lr){3-4}\cmidrule(lr){5-6}
Feature Representation / Weighting Arm & $D$ & Price FSS (\%) & IV Skill (\%) & Price FSS (\%) & IV Skill (\%) \\
\midrule
\multicolumn{6}{@{}l}{\textit{Panel A: Nested Feature \& Latent Representation Tiers ($\gamma_{\mathrm{HT}} = 0.55$; Left: Full Census, Right: PPS $\alpha=0.60$)}} \\
Tier 1: Price \& technicals only & 16 & $+0.17 \pm 0.10$ & $-0.1 \pm 0.2$ & $+0.06 \pm 0.13$ & $-0.3 \pm 0.3$ \\
Tier 2: + IV level, skew, $\Delta\mathrm{IV}_{1\mathrm{d}}$, VRP proxy & 20 & $-0.04 \pm 0.50$ & $+13.4 \pm 3.7$ & $-0.42 \pm 0.41$ & $+11.0 \pm 10.6$ \\
Tier 3: + 8 option chain activity aggregates & 28 & $-0.44 \pm 1.31$ & $+3.9 \pm 5.1$ & $-0.58 \pm 0.85$ & $+1.8 \pm 3.5$ \\
Tier 4a: + 3 PCA option-state latents (Augmentation) & 31 & $-0.26 \pm 0.58$ & $+7.7 \pm 3.1$ & $-1.08 \pm 0.38$ & $+7.7 \pm 1.1$ \\
Tier 4b: + 3 PCA option-state latents (Replacement) & 23 & $-0.07 \pm 0.21$ & $+13.0 \pm 5.0$ & $-0.40 \pm 0.69$ & $+12.3 \pm 5.3$ \\
Tier 5a: + 3 OptionVAE latents (Augmentation) & 31 & $-0.29 \pm 0.30$ & $+4.4 \pm 5.4$ & $-0.92 \pm 0.48$ & $+3.0 \pm 5.1$ \\
Tier 5b: + 3 OptionVAE latents (Replacement) & 23 & $+0.13 \pm 0.30$ & $+11.8 \pm 6.3$ & $-0.32 \pm 0.45$ & $+13.6 \pm 2.8$ \\
\midrule
\multicolumn{6}{@{}l}{\textit{Panel B: Weight-Tempering Sensitivity at $\alpha = 0.60$ (Left: Tier 5a Neyman / PPS; Right: Tier 5b Replacement PPS)}} \\
Unweighted subsampling ($\gamma_{\mathrm{HT}} = 0.00$, Tier 5a / 5b) & 31 / 23 & $-1.44 \pm 1.32$ & $+1.8 \pm 5.7$ & $-0.97 \pm 0.57$ & $+0.1 \pm 6.0$ \;($+12.5 \pm 4.1$) \\
Tempered Inverse-Probability Weighting ($\gamma_{\mathrm{HT}} = 0.55$, Tier 5a / 5b) & 31 / 23 & $-0.54 \pm 0.48$ & $+3.2 \pm 5.1$ & $-0.92 \pm 0.48$ & $+3.0 \pm 5.1$ \;($+13.6 \pm 2.8$) \\
Self-Normalized Inverse-Probability Weighting ($\gamma_{\mathrm{HT}} = 1.00$, H\'{a}jek Ratio IPW, Tier 5a / 5b) & 31 / 23 & $-0.35 \pm 0.44$ & $+3.2 \pm 4.0$ & $-0.60 \pm 0.32$ & $+5.5 \pm 5.2$ \;($+14.0 \pm 2.6$) \\
\bottomrule
\end{tabular}

%% file: tables/tab_bars.tex
\begin{tabular}{@{}lrrrcrcccr@{}}
\toprule
 & \multicolumn{3}{c}{Count \& Dispersion} & \multicolumn{4}{c}{Higher-Moment \& Non-Crossing ARCH Diagnostics} & \multicolumn{2}{c}{1-Bar LSTM} \\
\cmidrule(lr){2-4}\cmidrule(lr){5-8}\cmidrule(lr){9-10}
Clock & Return & Ret./ & Return & Excess Kurtosis & $\Delta\kappa$ vs.\ Time & $\Delta\kappa$ vs.\ Diurn. & $\rho_1(r_t^2)$ & $R^2_{\mathrm{OOS}}$ & DA \\
 & Count & sess. & SD (\%) & [95\% CI] & [95\% Boot. CI] & [95\% Boot. CI] & [95\% CI] & (\%) & (\%) \\
\midrule
\multicolumn{10}{@{}l}{\textit{Panel A: Interior-Session Close-to-Close Returns (Excluding Opening Bar, $j = 2, \dots, K_d$, $N = \mathrm{Bars} - 967$)}} \\
Time (Chronological) & 7,736 & 8.00 & 0.709 & 5.00 [3.20, 7.76] & -- & $-0.09$ $[-1.66, +1.13]$ & 0.187 [0.128, 0.232] & $-0.15$ & 48.6 \\
Time (Diurnally Std.) & 7,736 & 8.00 & 0.709 & 5.09 [2.48, 9.44] & $+0.09$ $[-1.13, +1.66]$ & -- & 0.216 [0.159, 0.264] & $-0.14$ & 48.7 \\
Volume (60d Causal) & 7,238 & 7.49 & 0.855 & 1.83 [1.05, 2.76] & $-3.17$ $[-5.99, -1.11]$ & $-3.26$ $[-6.88, -0.53]$ & 0.170 [0.129, 0.216] & $+0.05$ & 50.1 \\
Dollar (60d Causal) & 7,729 & 7.99 & 0.838 & 2.49 [1.74, 3.35] & $-2.51$ $[-4.86, -0.75]$ & $-2.60$ $[-5.66, -0.40]$ & 0.215 [0.164, 0.263] & $-0.18$ & 51.0 \\
\midrule
\multicolumn{10}{@{}l}{\textit{Panel B: Full-Session Intraday Returns ($09\text{:}30\text{--}16\text{:}00$ ET, Including Bar-1 Open-to-Close $r_{d,1} = \log(C_{d,1}/O_{d,1})$)}} \\
Time (Chronological) & 8,703 & 9.00 & 0.869 & 14.00 [4.86, 28.00] & -- & $+8.78$ $[-0.73, +22.58]$ & 0.157 [0.131, 0.199] & $-0.15$ & 48.6 \\
Time (Diurnally Std.) & 8,703 & 9.00 & 0.847 & 5.22 [2.60, 9.02] & $-8.78$ $[-22.58, +0.73]$ & -- & 0.206 [0.158, 0.249] & $-0.14$ & 48.7 \\
Volume (60d Causal) & 8,205 & 8.49 & 0.872 & 1.67 [1.04, 2.43] & $-12.33$ $[-25.97, -3.64]$ & $-3.55$ $[-7.40, -0.73]$ & 0.147 [0.109, 0.194] & $+0.05$ & 50.1 \\
Dollar (60d Causal) & 8,696 & 8.99 & 0.857 & 2.28 [1.67, 3.00] & $-11.72$ $[-25.30, -3.02]$ & $-2.94$ $[-6.28, -0.55]$ & 0.199 [0.154, 0.240] & $-0.18$ & 51.0 \\
\bottomrule
\end{tabular}

%% file: tab_bars.tex
\begin{tabular}{@{}lrrrcrcccr@{}}
\toprule
 & \multicolumn{3}{c}{Count \& Dispersion} & \multicolumn{4}{c}{Higher-Moment \& Non-Crossing ARCH Diagnostics} & \multicolumn{2}{c}{1-Bar LSTM} \\
\cmidrule(lr){2-4}\cmidrule(lr){5-8}\cmidrule(lr){9-10}
Clock & Return & Ret./ & Return & Excess Kurtosis & $\Delta\kappa$ vs.\ Time & $\Delta\kappa$ vs.\ Diurn. & $\rho_1(r_t^2)$ & $R^2_{\mathrm{OOS}}$ & DA \\
 & Count & sess. & SD (\%) & [95\% CI] & [95\% Boot. CI] & [95\% Boot. CI] & [95\% CI] & (\%) & (\%) \\
\midrule
\multicolumn{10}{@{}l}{\textit{Panel A: Interior-Session Close-to-Close Returns (Excluding Opening Bar, $j = 2, \dots, K_d$, $N = \mathrm{Bars} - 967$)}} \\
Time (Chronological) & 7,736 & 8.00 & 0.709 & 5.00 [3.20, 7.76] & -- & $-0.09$ $[-1.66, +1.13]$ & 0.187 [0.128, 0.232] & $-0.15$ & 48.6 \\
Time (Diurnally Std.) & 7,736 & 8.00 & 0.709 & 5.09 [2.48, 9.44] & $+0.09$ $[-1.13, +1.66]$ & -- & 0.216 [0.159, 0.264] & $-0.14$ & 48.7 \\
Volume (60d Causal) & 7,238 & 7.49 & 0.855 & 1.83 [1.05, 2.76] & $-3.17$ $[-5.99, -1.11]$ & $-3.26$ $[-6.88, -0.53]$ & 0.170 [0.129, 0.216] & $+0.05$ & 50.1 \\
Dollar (60d Causal) & 7,729 & 7.99 & 0.838 & 2.49 [1.74, 3.35] & $-2.51$ $[-4.86, -0.75]$ & $-2.60$ $[-5.66, -0.40]$ & 0.215 [0.164, 0.263] & $-0.18$ & 51.0 \\
\midrule
\multicolumn{10}{@{}l}{\textit{Panel B: Full-Session Intraday Returns ($09\text{:}30\text{--}16\text{:}00$ ET, Including Bar-1 Open-to-Close $r_{d,1} = \log(C_{d,1}/O_{d,1})$)}} \\
Time (Chronological) & 8,703 & 9.00 & 0.869 & 14.00 [4.86, 28.00] & -- & $+8.78$ $[-0.73, +22.58]$ & 0.157 [0.131, 0.199] & $-0.15$ & 48.6 \\
Time (Diurnally Std.) & 8,703 & 9.00 & 0.847 & 5.22 [2.60, 9.02] & $-8.78$ $[-22.58, +0.73]$ & -- & 0.206 [0.158, 0.249] & $-0.14$ & 48.7 \\
Volume (60d Causal) & 8,205 & 8.49 & 0.872 & 1.67 [1.04, 2.43] & $-12.33$ $[-25.97, -3.64]$ & $-3.55$ $[-7.40, -0.73]$ & 0.147 [0.109, 0.194] & $+0.05$ & 50.1 \\
Dollar (60d Causal) & 8,696 & 8.99 & 0.857 & 2.28 [1.67, 3.00] & $-11.72$ $[-25.30, -3.02]$ & $-2.94$ $[-6.28, -0.55]$ & 0.199 [0.154, 0.240] & $-0.18$ & 51.0 \\
\bottomrule
\end{tabular}